\documentclass[aps,prd,floatfix,nofootinbib,superscriptaddress,twocolumn,showkeys]{revtex4-2}

\usepackage{amsmath}

\usepackage{amssymb}
\allowdisplaybreaks

\usepackage{makeidx}
\usepackage{amsfonts}
\usepackage[usenames,dvipsnames]{pstricks}
\usepackage{subfigure}
\usepackage{epsfig}
\usepackage{pst-grad} 
\usepackage{mathtools}
\usepackage[colorlinks,hyperindex]{hyperref}
\usepackage{array}

\hypersetup
{	colorlinks,%
	citecolor=black,%
	linkcolor=black,%
	urlcolor=black,%
}

\usepackage{allpurpose}

\usepackage{flushend}

\newcommand\nn{{\nonumber}}

\begin{document}

\title{Deflection angle with electromagnetic interaction and Gravitational-electromagnetic dual lensing}

\author{Xiaoge Xu}
\address{School of Physics and Technology, Wuhan University, Wuhan, 430072, China}

\author{Tingyuan Jiang}
\address{School of Physics and Technology, Wuhan University, Wuhan, 430072, China}

\author{Junji Jia}
\email[Corresponding author:~]{junjijia@whu.edu.cn}
\address{Center for Astrophysics \& MOE Key Laboratory of Artificial Micro- and Nano-structures, School of Physics and Technology, Wuhan University, Wuhan, 430072, China}

\date{\today}

\begin{abstract}
The trajectory deflection and  gravitational-electromagnetic dual lensing (GEL) of charged signal in general charged static and spherically symmetric spacetimes are considered in this work. We showed that the perturbative approach previously developed for neutral particles can be extended to the electromagnetic interaction case. The deflection angle still takes a (quasi-)series form and the finite distance effect of both the source and observer can be taken into account. Comparing to pure gravitational case, the apparent angles of the images in the GEL, their magnifications and time delay all receive the electromagnetic corrections starting from the first non-trivial order. The sign and relative size of the leading corrections are determined by $\sim \frac{Q}{M}\frac{q}{E}$ where $M,~Q,~q,~E$ are the spacetime mass and charge, and signal particle charge and energy respectively. It is found that for $qQ>0$ (or $<0$), the electromagnetic interaction will decrease (or increase) the  deflection angle, and in GEL the impact parameters, apparent angles, magnifications and total travel time for each image. The time delay is increased for small $\beta$ and $qQ>0$, and otherwise always increased regardless the sign of $qQ$. The results are then applied to the deflection and GEL of charged protons in cosmic rays in Reissner-Nordstrom, charged     dilaton and charged Horndeski spacetimes. 

\end{abstract}


\maketitle

\section{Introduction}

Deflection of light in gravitational field was one of the most important supporting evidence in the development of General Relativity \cite{Dyson:1920cwa}. It is the foundation of gravitational lensing (GL), which itself has also become a very powerful tool in astrophysics and cosmology. GL can be used to investigate 
properties of the supernova \cite{Sharon:2014ija},
coevolution of galaxies and supermassive black holes (SMBHs) \cite{Peng:2006ew},
mass distributions of galaxy clusters \cite{Bartelmann:1999yn} and the Universe  \cite{Ade:2015zua},
cosmological parameters \cite{Refregier:2003ct,Lewis:2006fu} and dark matter and energy \cite{Metcalf:2001ap, Hoekstra:2008db}.

Besides the usual light rays, other types of signals such as neutrinos, gravitational waves (GW) and cosmic rays (CR) can also travel across large scale distance and become messengers of their sources,     carrying information about themselves and the spacetime they passed. With the discovery of SN 1987A neutrinos \cite{Hirata:1987hu, Bionta:1987qt} and the recent neutrinos from blazar TXS 0506+056 \cite{IceCube:2018dnn,IceCube:2018cha}, the more recent GWs \cite{Abbott:2016blz,Abbott:2016nmj,Abbott:2017oio,TheLIGOScientific:2017qsa} and multimessenger event \cite{Monitor:2017mdv}, neutrinos and GWs are now recognized as important tools in astrophysical observations. On the other hand, the CRs have an even longer observational history \cite{LetessierSelvon:2011dy},
and it has become clearer in recent years that CRs with ultra-high energies (UHE) are composed of charged massive particles of protons and  heavier nucleus \cite{AlvesBatista:2019tlv}. 

Previously, trajectory deflection and GL of neutral massive signals, including neutrinos and GWs in some gravitational theories beyond GR, have been studied by many researchers. The dependence of the deflection angle and GL observables on the general metric functions and signal properties have been studied in the weak field limit using both the perturbative method  \cite{Jia:2015zon, He:2016vxc, Pang:2018jpm, Huang:2020trl, Liu:2020wcu, Liu:2020mkf,Duan:2020tsq, Jia:2020xbc,He:2020eah} and Gauss-Bonnet theorem method \cite{Crisnejo:2018uyn,Jusufi:2018kry,Jusufi:2019rcw,Li:2019vhp,Crisnejo:2019ril,Javed:2019ynm,Li:2019qyb,Li:2020dln}. Recently, there is a resurgence of interest to study the deflection of charged massive particles (CMP) in charged spacetimes \cite{Crisnejo:2019xtp,Jusufi:2019rcw,Li:2020ozr}. Although there might be intergalactic magnetic field that can cause chaotic motions of these signals, not all intergalactic space carry (strong) magnetic field. Particularly, for CMP with UHE, deflection due to extragalactic magnetic field is expected to be small \cite{AlvesBatista:2019tlv}. 

In this work we will study the deflection and the dual gravitational and electromagnetic lensing (GEL) of CMP in general charged  static and spherical symmetric (SSS) spacetimes. We will investigate whether the perturbative method that was developed previously for neutral particles in general SSS or stationary axisymmetric spacetimes in the weak field limit \cite{Huang:2020trl, Liu:2020wcu, Liu:2020mkf,Duan:2020tsq, Jia:2020xbc} can be extended to the case with electromagnetic interaction. It turns out that the extension is possible with only a small modification of the derivation process (see Sec. \ref{sec:pertmeth}). Using this method we will compute the deflection angle of CMP in general charged SSS spacetime with finite distance effect of the source and detector and the electromagnetic interaction taken into account. The image apparent angles and time delay in the GEL, including the leading order contribution from the electromagnetic interaction, will also be given (Sec. \ref{sec:gltd}). We then apply the results to the Reissner-Nordstrom, Gibbons-Maeda-Garfinkle-Horowitz-Strominger (GMGHS) \cite{Gibbons:1982ih,Gibbons:1987ps,Garfinkle:1990qj}
and charged Horndeski \cite{Cisterna:2014nua} spacetimes to verify their validity and show more clearly the effect of the electromagnetic interaction on various quantities (Sec. \ref{ssec:apptoqs}). We use the natural units $G=c=1/(4\pi\epsilon_0)=1$ in this work.

\section{The perturbative method\label{sec:pertmeth}}

We start from the SSS metric with line element
\begin{equation}
\dd s^2=-A(r) \dd t^2+B(r)\dd r^2+C(r) (\dd\theta^2+\sin^2\theta\dd\phi^2) \label{eq:sssmetric}
\end{equation}
where $(t,~r,~\theta,~\phi)$ are the coordinates and $A(r),~B(r)$ and $C(r)$ are the metric functions. We assume that the spacetime only possesses a known SSS electric field but no magnetic field. Therefore the field can be described by the four-potential $A_{\mu}(r)$ with $A_0\neq0, ~A_i=0 ~(i=1,2,3)$.

The motion of charged particle with mass $m$ and charge $q$  under both gravity and electric interaction is described by the Lorentz equation~\cite{Rohrlich}
\begin{align}
\frac{d^2{x}^{\rho}}{d\tau^2}+\Gamma^{\rho}_{\mu\nu}\frac{dx^{\mu}}{d\tau}\frac{dx^{\nu}}{d\tau}= \frac{q}{m} F^{\rho}_{\mu}\frac{d{x}^{\mu}}{d\tau}~, \label{eq:lorentzeq}
\end{align}
where $\tau$ is the proper time or affine parameter of the particle, and $F_{\mu\nu}=\partial_{\mu}A_{\nu}-\partial_{\nu}A_{\mu}$ is the electromagnetic field strength. From this we can see that the motion of the particle no longer follows a geodesic in the background spacetime.

Previously,  for neutral particles, we have computed the deflection angle using a perturbative method developed in Ref. \cite{Huang:2020trl,Liu:2020mkf}. It turns out this method can be extended to the current case of charged particles under dual influence of gravity and electric interactions, with a small modification at essentially only one place in the equations. Here we will briefly recap the perturbative method and point out this modification, and show that this method is applicable to all SSS spacetime with arbitrary electric potential $A_0(r)$.

First of all,  the time, angular and radial components of Eq. \eqref{eq:lorentzeq}, after first integrals, become
\bea
&&\dot{t}=\frac{E+qA_0}{mA},\label{eq:teq}\\
&&\dot{\phi}=\frac{L}{mC}, \label{eq:phieq}\\
&&\dot{r}^2=\frac{\lsb (E+qA_0)^2-m^2 A\rsb C-L^2A}{m^2ABC},\label{eq:req}
\eea
where without losing any generality we have set the trajectory to be in the equator plane, i.e., $\theta=\pi/2$. Here the $L$ and $E$ are constants of first integrals allowed by the SSS symmetry and can be interpreted as the angular momentum and energy of the particle. $L$ can be related to the trajectory's minimal radius $r_0$  using Eq. \eqref{eq:req}, i.e. $\dot{r}|_{r=r_0}=0$, to find 
\be
L=\sqrt{[(E+qA_0)^2-m^2 A(r_0)]C(r_0)/A(r_0)}. \label{eq:linr0}
\ee
Comparing to the case without the electric interaction \cite{Huang:2020trl}, the effect of the charge manifests in Eqs. \eqref{eq:teq}-\eqref{eq:linr0} through the terms involving $qA_0$. In asymptotically flat spacetimes (on which this work concentrates), 
$L$ and $E$ can also be related to the signal's impact parameter $b$ and asymptotic velocity $v$ by 
\be
L=|\vecr\times \vp|=\frac{mv}{\sqrt{1-v^2}}b,~E=\frac{m}{\sqrt{1-v^2}}. \label{eq:leandbv}
\ee

From Eqs. \eqref{eq:teq} to \eqref{eq:req}, the change of the angular coordinate $\Delta\phi$ from a source located at $(r_s,\phi_s)$ to a detector located at $(r_d,\phi_d)$ (see Fig. \ref{fig:glcharge}) becomes
\be
\Delta\phi=\lsb \int_{r_0}^{r_s}+\int_{r_0}^{r_d}\rsb \sqrt{\frac{B}{C}}\frac{L}{\sqrt{\lsb (E+qA_0)^2/A-m^2 \rsb C-L^2}}\dd r,
\label{eq:deltaphi}
\ee
while the total flight time $\Delta t$  becomes
\be \Delta t=\lsb \int_{r_0}^{r_s}+\int_{r_0}^{r_d}\rsb \frac{\sqrt{BC}}{A}\frac{E+qA_0}{\sqrt{\lsb (E+qA_0)^2/A-m^2\rsb C-L^2}}\dd r.\label{eq:deltat}
\ee
The integrations in Eqs. \eqref{eq:deltaphi} and \eqref{eq:deltat} usually can not be explicitly carried out for general SSS  metrics and therefore require a perturbative technique to find their approximations.

To proceed, one can replace the $L,~E$ and $r_0$ in Eqs.  \eqref{eq:deltaphi} and \eqref{eq:deltat} by $b$ and $v$. Solving $b$ from Eq. \eqref{eq:leandbv} and using Eq. \eqref{eq:linr0}, we have
\bea
\frac{1}{b}&=&\frac{\sqrt{E^2-m^2}}{\sqrt{(E+qA_0(r_0))^2-m^2 A(r_0)}}\sqrt{\frac{A(r_0)}{C(r_0)}}\label{eq:pfuncdefqunt}\\
&\equiv& p\lb \frac{1}{r_0}\rb. \label{eq:pfuncdef}
\eea
Here we have defined the right-hand side of Eq. \eqref{eq:pfuncdefqunt} as a function $p$ of $1/r_0$. For later usage, we will denote the inverse function of $p(x)$ as $w(x)$. We emphasize that once the metric functions and four-potential are known, the function $p(x)$ will be explicitly known too. Although obtaining the analytical form of a function's inverse function, in this case $w(x)$, can sometimes be difficult, its series form can always be obtained using the Lagrange inversion theorem. And this series form of $w(x)$ is all we need for later perturbative computations. 

One of the difficulties in the integration of Eqs. \eqref{eq:deltaphi} and \eqref{eq:deltat} comes from the fact that the minimal radius $r_0$, which is usually difficult to link to observables but often used as an expansion parameter, appears in the lower limit. To solve this problem, in Ref. \cite{Huang:2020trl,Liu:2020mkf}, we have proposed the change of variables from $r$ to $u$ which are linked by $w(x)$ (but without particle charge there). In the current situation, it appears that a similar change of variables from $r$ to $u$ can still be used as long as we use the updated $p(x)$ and $w(x)$ given in Eq. \eqref{eq:pfuncdef}, i.e., 
\be
\frac{1}{r}=w\lb \frac{u}{b}\rb~\text{or equivalently}~r=1/w\lb \frac{u}{b}\rb . \label{eq:rtoucov}
\ee
With this and also using Eqs. \eqref{eq:linr0} and \eqref{eq:pfuncdef}, the integral limits and various terms in the integrand of Eqs. \eqref{eq:deltaphi} and \eqref{eq:deltat}  can be transformed to
(also see \cite{Liu:2020mkf})
\begin{subequations} \label{eq:covtrans}
\begin{align}
&r_0\to 1,~r_{s,d}\to b\cdot p\lb\frac{1}{r_{s,d}}\rb ,\\
&\dd r\to  -\frac{1}{p^\prime(w)w^2}\frac1{b}\dd u,\\
&
\sqrt{\frac{B}{C}}\to \sqrt{\frac{B(1/w)}{C(1/w)}},\\
&\frac{L}{\sqrt{\lsb (E+qA_0)^2/A-m^2 \rsb C-L^2}}\to \frac{u}{\sqrt{1-u^2}}.
\end{align}
\end{subequations}
Combining right-hand side of Eq. \eqref{eq:covtrans}, the $\Delta\phi$ becomes
\be
\Delta\phi=\lsb \int_{\sin\theta_s}^{1}+\int_{\sin\theta_d}^{1}\rsb y\lb \frac{u}{b} \rb \frac{\dd u}{\sqrt{1-u^2}} \label{eq:phiinu}
\ee
where we have collectively denoted part of the integrand as
\be
y\lb \frac{u}{b}\rb =\sqrt{\frac{B(1/w)}{C(1/w)}}\frac{1}{p'(w)w^2}\frac{u}{b},~~\text{with}~~w=w\lb \frac{u}{b}\rb \label{eq:yubdef}
\ee
Note in Eq. \eqref{eq:phiinu} 
\be
\theta_s=\arcsin\lsb b\cdot p\lb \frac{1}{r_s}\rb \rsb,~\theta_d=\arcsin\lsb b\cdot p\lb \frac{1}{r_d}\rb \rsb
\label{eq:appang}
\ee
are respectively the apparent angles of the signal at the source and detector \cite{Huang:2020trl}.
Similar to $\Delta\phi$, the total travel time $\Delta t$ in Eq. \eqref{eq:deltat} becomes
\be
\Delta t=\lsb \int_{\sin\theta_s}^{1}+\int_{\sin\theta_d}^{1}\rsb z\lb \frac{u}{b} \rb \frac{\dd u}{u\sqrt{1-u^2}} \label{eq:tinu}
\ee
where
\bea z\lb \frac{u}{b}\rb &=&
\frac{\sqrt{B(1/w)C(1/w)}}{A(1/w)} \frac{u}{b}\frac{1}{v}
\frac{1}{p^\prime ( w) w^2}
\frac{u}{b}\nn\\
&&\times\lsb 1+\hat{q}A_0(1/w)\sqrt{1-v^2}\rsb \label{eq:zdef}
\eea
and $\hat{q}\equiv q/m$ is the charge-mass ratio of the signal particle.

Now we restrict ourselves to the weak field limit, in which the impact parameter is much larger than the characteristic mass of the spacetime. Therefore in this limit we can expand respectively the $y\lb \frac{u}{b}\rb $ factor in Eq. \eqref{eq:phiinu} and $z\lb \frac{u}{b}\rb $ factor in Eq. \eqref{eq:tinu} into series of $\frac{u}{b}$
\bea
&&y\lb \frac{u}{b}\rb =\sum_{n=0} y_n \lb  \frac{u}{b}\rb^n, \label{eq:yexp}\\
&&z\lb \frac{u}{b}\rb =\sum_{n=-1}z_n \lb  \frac{u}{b}\rb^n, \label{eq:zexp}
\eea
where the two series start respectively from $n=0$ and $n=-1$ because of the form of the metric functions and electric potential given in Eq. \eqref{eq:abca0exp}. 

The important point is that these coefficients $y_n$ in Eq. \eqref{eq:ynexp} and $z_n$ in Eq. \eqref{eq:znexp} can always be expressed in terms of the coefficients of the asymptotic expansions of the metric functions and the electric potential, which we will assume to be 
\bea
&& A(r)=\sum_{n=0}^\infty\frac{a_n}{r^n},~
B(r)=\sum_{n=0}^\infty\frac{b_n}{r^n},~
C(r)=r^2\sum_{n=0}^\infty\frac{c_n}{r^n},\nn\\
&&
A_0(r)=\sum_{n=1}^\infty\frac{a_{0n}}{r^n}.
\label{eq:abca0exp}\eea
Without losing any generality, we can always choose $a_0=b_0=c_0=1$ and identify $a_1=-2M,~a_{01}=-Q$ where $M$ and $Q$ are respectively the ADM mass and total charge of the spacetime. In principle, we can even set $C(r)=r^2$ exactly because the allowed change of the coordinates. However for the moment we will allow general $c_n~(n\geq 1)$ because some works presented their $C(r)$ in other forms. 
Using the expansions \eqref{eq:abca0exp} in Eq. \eqref{eq:pfuncdef}, we can obtain the series form of function $p(x)$ as
\be 
p(x)=x+\lb \frac{a_1}{2v^2}-\frac{c_1}{2}-\frac{a_{01}\hat{q}\sqrt{1-v^2}}{v^2}\rb x^2+\mathcal{O}(x)^3. \label{eq:pexp}
\ee
Inverting this series, we can obtain a series form of $w(x)$ and then substituting it together with Eq. \eqref{eq:abca0exp} into Eqs. \eqref{eq:yubdef} and \eqref{eq:zdef}, we can further expand them according to Eqs. \eqref{eq:ynexp} and \eqref{eq:znexp} and the first few coefficients are found to be
\begin{subequations}
\label{eq:ynexp}
\begin{align}
y_0=&1,\\
y_1=&-\frac{a_1}{2 v^2}+\frac{b_1}{2}+\frac{a_{01} \hat{q} \sqrt{1-v^2}}{v^2},\label{eq:y1exp}\\
\nonumber y_2=&-\frac{1}{2 v^2}\lsb a_1\lb  b_1+c_1-2a_1\rb+2a_2\rsb\\
&-\frac{1}{8}\lb b_1+c_1\rb^2+\frac{1}{2}\lb b_2+c_2\rb \nn \\
&+\lsb (b_1+c_1-2a_1) a_{01}+2a_{02}\rsb \frac{ \hat{q} \sqrt{1-v^2}}{v^2}\nn\\
&+\lb \frac{1}{ v^2}-1\rb a_{01}^2 \hat{q}^2,
\end{align}
\end{subequations}
 and
\begin{subequations}
\label{eq:znexp}
\begin{align}
z_{-1}=&\frac{1}{v},\\
 z_0=&\frac{1}{v}\lb -a_1+\frac{b_1}{2 }+\frac{a_1}{2 v^2}\rb +\lb 1-\frac{1}{v^2}\rb \frac{a_{01} \hat{q} \sqrt{1-v^2}}{v} ,\\
\nonumber z_1=&\frac{1}{2 v} \lsb a_1 (2 a_1-b_1-c_1)-2 a_2+b_2+c_2\rsb-\frac{1}{8 v}\lb b_1-c_1\rb^2\\
&+\lsb \lb b_1+c_1-2 a_1\rb +2a_{02 }\rsb \frac{a_{01} \hat{q} \sqrt{1-v^2}}{2 v}.
\end{align}
\end{subequations}
Higher-order coefficients can also be easily obtained but they are only shown in Appendix \ref{sec:appdyzhigh} for their excessive length. 

A few comments are in order here. Firstly, setting $q=0$ in Eqs. \eqref{eq:ynexp} and \eqref{eq:znexp}, the $y_n$ and $z_n$ reduce to the case of neutral particles (without electromagnetic interaction), which  agree with results in  Ref. \cite{Huang:2020trl, Liu:2020mkf}. Secondly, the electromagnetic interaction affects both $\Delta \phi$ and $\Delta t$ from the first non-trivial orders ($n=1$ and $n=0$ respectively) through $a_{01}$, in contrast to the gravitational effect of spacetime charge which only appears from the second non-trivial order through $a_2$ or $b_2$ (see Eq. \eqref{eq:rncoeff} for their values in the RN spacetime). Thirdly, we see that all terms involving the potential $A_0(r)$ or equivalently $a_{0n}$ are multiplied by positive powers of $q$, and vice versa (also see Eq. \eqref{eq:y3general}). This implies that the electromagnetic interaction between the particle and central lens contributes to the deflection angle or flight time only through these terms. Furthermore,  from Eqs. \eqref{eq:ynexp} and \eqref{eq:znexp} one observes that quantitatively the electromagnetic interaction correction is proportional to $\sim \hat{q}\sqrt{1-v^2}=\frac{q}{m}\cdot\frac{m}{E}=q/E$, i.e, the inverse of the signals' {\it rigidity}. Although for a relativistic particle its $m/E$ is small and the charge of spacetime is also expected to be (much) smaller than its mass, the charge-mass ratio $q/m$, when converted to a dimensionless quantity, is actually very large, being $1.11\times 10^{18}$ for protons and $2.04\times 10^{21}$ for electrons. Therefore the electromagnetic correction terms are not necessarily small comparing to other terms in the coefficients. This is a reflection of the very fact that the electromagnetic interaction strength is much stronger than that of the gravitational interaction. Because of this possible largeness and the fact that we are carrying out a {\it perturbative} computation, the validity of the method developed in this work requires the condition that 
\be 
\mathcal{O}\lb a_{01}\hat{q}\sqrt{1-v^2}\rb \approx \mathcal{O}\lb \frac{a_{01}q}{E}\rb <\mathcal{O}(b). \label{eq:pertvalidcond}
\ee
Finally, the $\Delta\phi$ of massless charged particles can be obtained by systematically replacing $\hat{q}\sqrt{1-v^2}\to q/E$ and then taking the $v\to 1$ limit in Eq. \eqref{eq:yexp}. 

Substituting Eqs. \eqref{eq:ynexp} and \eqref{eq:znexp} into Eqs. \eqref{eq:phiinu} and \eqref{eq:tinu}, $\Delta\phi$ and $\Delta t$ both become sums of a series of integrals of the form 
\be
I_n(\theta_i)= \int_{\sin\theta_i}^{1}
\frac{u^n}{\sqrt{1-u^2}}\dd u
\ee
which can always be integrated into elementary functions and their explicit forms are given in Eq. \eqref{eq:inthetares} in Appendix \ref{sec:appd}. 
Consequently we found an effective way to approximate the initial  $\Delta\phi$ and $\Delta t$
\bea
&&\Delta\phi =\sum_{i=s,d}\sum_{n=0}^\infty y_n\frac{I_n(\theta_i)}{b^n} , \label{eq:dphifinal}\\
&&\Delta t =\sum_{i=s,d}\sum_{n=-1}^\infty z_n\frac{I_{n-1}(\theta_i)}{b^n} . \label{eq:dtfinal}
\eea

For the change of the angular coordinate, to the first three orders, Eq. \eqref{eq:dphifinal} is
\begin{align} 
\Delta\phi
=&\sum_{i=s,d}\lcb \lb \frac{\pi}{2}-\theta_i\rb y_0 +c_i\frac{y_1}{b} +\frac14(\pi +2s_ic_i-2\theta_i)\frac{y_2}{b^2}\right.\nn\\
&\left.+\mathcal{O}\lb \frac{M}{b}\rb^3\rcb, \label{eq:dphifinal3}\end{align}
where $y_n~(n=0,1,2)$ are given in Eq. \eqref{eq:ynexp} and $s_i=\sin\theta_i$ and $c_i=\cos\theta_i~(i=s,d)$. It is seen that the trivial leading order result with $y_0=1$ is given by $\pi -\theta_s-\theta_d$, which corresponds to the straight trajectory result. For the first non-trivial order result, i.e. $y_1/b$, using Eq. \eqref{eq:y1exp}, it is seen that comparing to the neutral signal case, $\Delta\phi$ receives a correction 
from the electromagnetic interaction, which as pointed out previously is not necessarily small. 
The finite distance effect of the source and detector manifests through terms involving $\theta_i~(i=s,~d)$ defined in Eq. \eqref{eq:appang}. 
To see this effect more clearly, we can expand $I_n$ in the small $b/r_i$ limit. Using the first few of these expansions given in Eq. \eqref{eq:ifirstthreeexp}, $\Delta\phi$ is further transformed to
\begin{align} 
\Delta\phi
=&\sum_{i=s,d}\lcb \frac{\pi}{2} y_0 + \frac{y_1}{b} +\frac{\pi}{4}\frac{y_2}{b^2}-\frac{b}{r_i}y_0\right.\nn\\
&-\frac{b}{r_i^2}\lsb \lb\frac{a_1}{2v^2}-\frac{c_1}{2}-\frac{a_{01}\hat{q}\sqrt{1-v^2}}{v^2}\rb y_0+y_1\rsb \nn\\
&\left. 
+\mathcal{O}\lb \frac{M}{b}\rb^3\rcb, \label{eq:dphifinal3exp}
\end{align}
The first three terms of this result is the infinite distance deflection angle with the electromagnetic interaction taken into account. 

\section{GEL and time delay\label{sec:gltd}}

Now with results \eqref{eq:dphifinal} and \eqref{eq:dtfinal}, in principle we can simply obtain $\Delta\phi$ and $\Delta t$ for any specific metric functions and four-potential. Before going to specific spacetimes however, in this section we will set up the GEL equations in general charged SSS spacetimes to solve the apparent angles of images and the time delay between them.

\begin{figure}[htp!]
    \centering
    \includegraphics[width=0.45\textwidth]{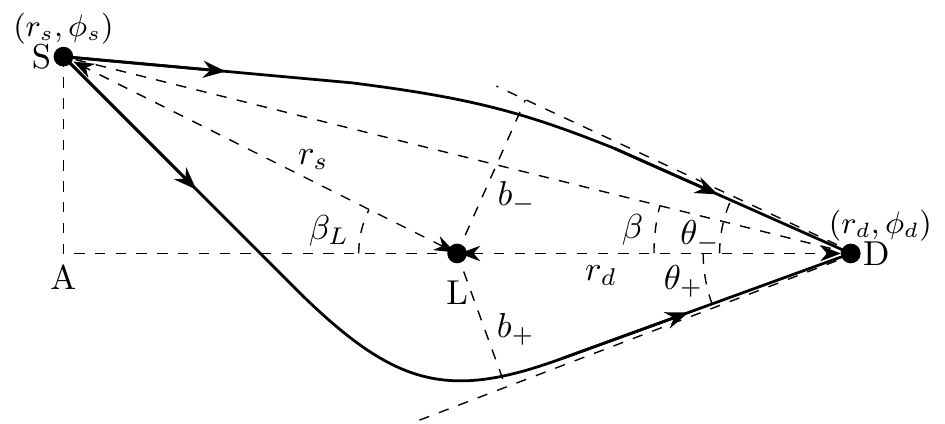}
    \caption{The GEL setup. The source is located at angle $\beta_L$ and $\beta$ respectively with respect to the lens and detector against the lens-detector axis. The impact parameters and the corresponding apparent angles corresponding to counter-clockwise and clockwise trajectories are labeled as $b_+$ and $\theta_+$, and $b_-$ and $\theta_-$ respectively.}
    \label{fig:glcharge}
\end{figure}

The setup of the GEL is illustrated in Fig. \ref{fig:glcharge}. Since our change of the angular coordinate \eqref{eq:dphifinal} is applicable to source and detectors at finite distance, we can use it to establish an exact GEL equation linking the source angular position $\beta_L$ and two impact parameters $b_\pm$ from each side of the lens. With $b_\pm$ solved from this equation, then using Eq. \eqref{eq:appang} we can find the corresponding apparent angles of the images, after which their magnifications and furthermore the time delays between the two signals can also be solved. 

The exact GEL equation is nothing but the following relation, or equivalently the very definition of $\Delta\phi$
\be\Delta\phi(b_\pm)=\pi\pm\beta_L\ee
where the $+$ and $-$ signs correspond to rays propagating counter-clockwise and clockwise respectively (see Fig. \ref{fig:glcharge}). Substituting $\Delta\phi$ from Eq. \eqref{eq:dphifinal3exp}
truncated to the first non-trivial orders (i.e. $(M/b)^1$ and $(b/r_i)^1$) into the above equation, it becomes \be 
\pi +\lb b_1- \frac{a_1}{v^2}+\frac{2a_{01}\hat{q}\sqrt{1-v^2}}{v^2}\rb \frac{1}{b} -b \lb \frac{1}{r_s}+\frac{1}{r_d}\rb 
=\pi \pm \beta_L. \ee
From this, one can solve the two impact parameters 
\begin{align}
b_\pm=&\frac{\beta_L r_d r_s }{2 (r_d+r_s)} \lsb \sqrt{1+\eta}\mp 1\rsb
\label{eq:blrtophi}
\end{align}
where 
\begin{align}
\eta
=\frac{8  (r_d+r_s)}{\beta_L^2 r_d r_s}\lb-\frac{a_1}{2 v^2}+\frac{b_1}{2}+\frac{a_{01} \hat{q} \sqrt{1-v^2}}{v^2}\rb .
\label{eq:etainbetal}
\end{align} 
Comparing to the corresponding result of neutral particles (Eq. (30) of Ref. \cite{Liu:2020wcu}), the electromagnetic interaction term $a_{01}\hat{q}$ now has a non-trivial contribution to $b_\pm$. Since $a_1=-2M$ and $a_{01}=-Q$, we see from Eq. \eqref{eq:etainbetal} that if $\mathrm{sign}(qQ)=1$ (or $-1$), i.e., the electromagnetic interaction between the particle and the lens is repulsive (or attractive), then $\eta$ will be smaller (or larger) than the neutral particle case. Using Eq. \eqref{eq:blrtophi}, this implies that electric repulsion (or attraction) between the signal and lens charges will force the desired impact parameters from both sides to decrease (or increase). 

\begin{figure}[htp!]
    \centering
    \includegraphics[width=0.45\textwidth]{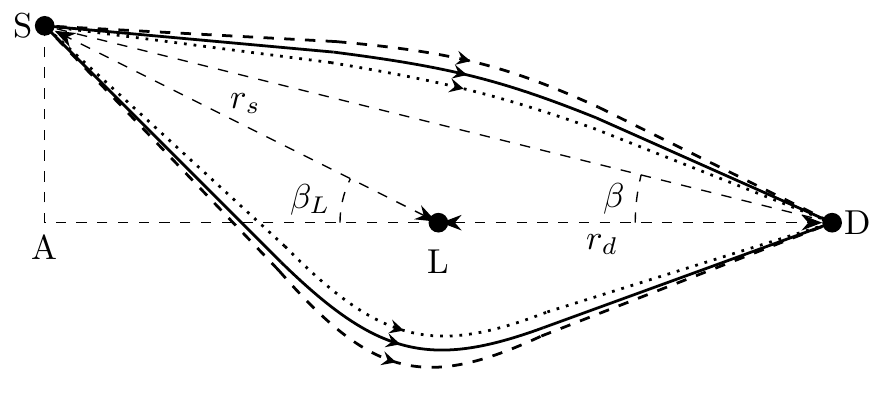}
    \caption{The GEL setup for signal with $\mathrm{sign}(qQ)=-1$ (dashed), neutral (solid) and $\mathrm{sign}(qQ)=1$ (dotted). The impact parameters, apparent angles of the images and Einstein radius will all be larger (or smaller) for signal with $\mathrm{sign}(qQ)=-1$ (or $1$) comparing to neutral particles.}
    \label{fig:gel2}
\end{figure}

For the apparent angles of the images, from Eq. \eqref{eq:appang} and using \eqref{eq:pexp}, to the second order of $1/r_d$, we have
\be 
\theta_\pm =b_\pm \lsb \frac{1}{r_d}
+\lb \frac{a_1}{2v^2}-\frac{c_1}{2}-\frac{a_{01}\hat{q}\sqrt{1-v^2}}{v^2}\rb \frac{1}{r_d^2}\rsb 
\label{eq:thetapmres}
\ee
where $b_\pm$ is given by Eq. \eqref{eq:blrtophi}. For $\beta_L=0=\beta$, the corresponding Einstein ring has an angular size 
\be 
\theta_\mathrm{E}=\sqrt{\frac{2r_s}{r_d(r_s+r_d)}\lb-\frac{a_1}{2 v^2}+\frac{b_1}{2}+\frac{a_{01} \hat{q} \sqrt{1-v^2}}{v^2}\rb}.
\label{eq:thetaeres}
\ee
Since the second-order terms of $\theta_\pm$ is much smaller than the leading $1/r_d$ order term, the apparent angles $\theta_\pm$ and the corresponding $\theta_\mathrm{E}$ will both be affected by the electromagnetic interaction in the same manner as $b_\pm$ were affected. That is, $\mathrm{sign}(qQ)=1$ (or $-1$) will lead to smaller (or larger) apparent angle of the images. The deviation of the trajectories of charges with different signs from that of neutral particles, and the change of the corresponding $b_\pm,~\theta_\pm$, are qualitatively represented in Fig. \ref{fig:gel2} by the dashed ($\mathrm{sign}(qQ)=-1$) and dotted  ($\mathrm{sign}(qQ)=1$) lines. 

To compute the magnifications of the images, which are defined as
\be
\mu_\pm=\left|\frac{\theta_\pm}{\beta}\frac{\dd \theta_\pm}{\dd \beta}\right|, 
\label{eq:mupmdef}
\ee
we should link the apparent angle $\theta_\pm$ with $\beta$ (see Fig. \ref{fig:glcharge}). This can be realized by the geometrical relation 
in triangles $\triangle SAL$ and $\triangle SAD$
\be 
r_s\sin\beta_L=\lb r_d+r_s\cos\beta_L\rb \tan\beta. 
\label{eq:betalbetarel}
\ee
Therefore after substituting Eq. \eqref{eq:thetapmres} into Eq. \eqref{eq:mupmdef} and using Eq. \eqref{eq:betalbetarel}, the magnifications become
\be
\mu_\pm=\left|\frac{\theta_\pm}{\beta}\frac{\dd \theta_\pm}{\dd \beta_L}\frac{\dd \beta_L}{\dd \beta}\right|=\frac{u_\beta^2+2}{2 u_\beta \sqrt{u_\beta^2+4}}\mp\frac{1}{2} \label{eq:magnif}
\ee
where $u_\beta=\beta/\theta_\mathrm{E}$.
Formally this takes the same form as the magnification in simple Schwarzschild BH lensing \cite{Proceedings:2006uym}, but note the $\theta_\mathrm{E}$ is now given by Eq. \eqref{eq:thetaeres}, containing electromagnetic interaction. Clearly, a larger $\theta_\mathrm{E}$ (i.e. $\mathrm{sign}(qQ)=-1$) causes a smaller $u_\beta$, which in turn causes the magnifications of both images to become larger than the neutral signal case. Similarly, for charges with $\mathrm{sign}(qQ)=1$, both magnifications will be smaller than the neutral case.

With the two impact parameters $b_\pm$ known in Eq. \eqref{eq:blrtophi}, using the travel time result \eqref{eq:dtfinal} with the first three $I_n~(n=-1,0,1)$ given in Eq. \eqref{eq:ifirstthreeexp}, one can subtract the two travel times $\Delta t(b_+)$ and  $\Delta t(b_-)$  from each other to obtain the time delay between the two images. To the leading orders of $M/b$ and $b/r_i$, the result is found to be
\bea 
&&\Delta^2 t_\pm\equiv \Delta t(b_+)-\Delta t(b_-)\nn\\
&=&
\left(-a_1 +b_1 v^2 +2 a_{01} \hat{q} \sqrt{1-v^2}\right)\frac{2 \sqrt{1+\eta}}{ v^3 \eta}\nn\\
&&+
\lsb \lb\frac{2}{v}-\frac{1}{v^3}\rb a_1
-\frac{b_1}{v} +2a_{01} \hat{q} \left(\frac{1}{v^2}-1\right)^{\frac{3}{2}}\rsb\nn\\ &&\times  
 \ln \frac{\sqrt{1+\eta}-1}{\sqrt{1+\eta}+1}
+\mathcal{O}\lb \beta M,\frac{b^3}{r_{s,d}^2},\frac{M^2}{b}\rb 
\label{eq:dtleading}
\eea
where $\eta$ is given by Eq. \eqref{eq:etainbetal} or in terms of $\beta$ as
\begin{align}
\eta(\beta)
=\frac{8 r_s}{\beta^2  (r_d+r_s)r_d }\lb-\frac{a_1}{2 v^2}+\frac{b_1}{2}+\frac{a_{01} \hat{q} \sqrt{1-v^2}}{v^2}\rb .
\label{eq:etainbeta}
\end{align} 
Formally, Eq. \eqref{eq:dtleading} agrees with the time delay for neutral particles (see Eq. (32) of Ref. \cite{Liu:2020wcu}) except now the $\eta$ contains contribution from electromagnetic interaction. In the limit 
\be \eta\gg 1, ~\text{i.e.}~M/r_{d,s}\gg \beta_L^2\sim\beta^2, 
\label{eq:largeetacond}
\ee
the two terms in Eq. \eqref{eq:dtleading} contribute similarly to the total time delay, and the result can be further simplified by the large $\eta$ expansion to be
\bea 
\Delta^2 t_\pm&=&\frac{4\lsb (b_1-a_1)+a_{01}\hat{q}\sqrt{1-v^2}\rsb}{v\sqrt{\eta}} \nn\\
&&+
\mathcal{O}\lsb M\lb \frac{r_{d,s}}{M}\beta^2\rb^{\frac{3}{2}}, \beta M, \frac{b^3}{r_{s,d}^2}, \frac{M^2}{b}\rsb .\label{eq:dtetalarge}
\eea 
In the opposite limit, that is 
\be \eta\ll 1,~M /r_{s,d}\ll \beta_L^2\sim\beta^2,\label{eq:largebetacond}
\ee
the time delay becomes 
\be
\Delta^2 t_\pm=\frac{2\lb b_1v^2-a_1+2a_{01}\hat{q}\sqrt{1-v^2}\rb}{v^3\eta} +
\mathcal{O}\lb M, \frac{b^3}{r_{s,d}^2}\rb .\label{eq:dtetasmall}
\ee
The effect of electromagnetic interaction on the time delay is slightly more tricky to analyze than that on the impact parameters and apparent angles of the images. 
For both Eqs. \eqref{eq:dtetalarge}
and \eqref{eq:dtetasmall}, we see that at the lowest order, both the numerator and denominator will deviate from the neutral particle value in the same direction as $\hat{q}$ deviates from zero (see Eq. \eqref{eq:etainbetal} for $\eta$). Therefore we will have to go to one order higher to figure out the effect of the charge on $\Delta^2 t_\pm$. A small perturbative expansion shows that the deviation of the time delay $\Delta^2 t_\pm$ from the neutral particle value is given by $\tilde{c} \hat{q}Q$ with $\tilde{c}>0$ when $|qQ|$ and $\beta$ are both small (see Eq. \eqref{eq:tdsmallbeta} for its value in the RN spacetime). Therefore for small $\beta$, in contrast to $b_\pm,~\theta_\pm$ and $\mu_\pm$, when $\mathrm{sign}(qQ)=1$ (or $-1$), $\Delta^2 t_\pm$ will be larger (or smaller) comparing to the neutral particle case. For larger $|qQ|$ and $\beta$, as we will see in Sec. \ref{subsec:rnres}, the time delay between charged signals of both signs will be larger than that of neutral signals.

In summary, let us make the observation from Eqs. \eqref{eq:y1exp}, \eqref{eq:dphifinal3exp}, \eqref{eq:blrtophi} with \eqref{eq:etainbeta},  \eqref{eq:thetapmres}, \eqref{eq:mupmdef} with \eqref{eq:thetaeres}, that the electromagnetic corrections to the gravitational values of all these quantities studied in this work depend on the relative size of the following quantities
\be 
\mathcal{O}(M),~\mathcal{O}\lb Qq/E\rb. \label{eq:ordercomp}
\ee
Here for the order estimation purpose we have limited ourselves to relativistic signals so that $v$ is close to 1, and without losing any generality we used $\mathcal{O}(b_1-a_1)\approx \mathcal{O}(M)$ and $\mathcal{O}(-a_{01})\approx \mathcal{O}(Q)$. 
If the latter in Eq. \eqref{eq:ordercomp} is smaller than the former, then the gravitational contributions to these quantities are still larger, and otherwise the electromagnetic terms become dominant.

\section{Application to the charged spacetimes\label{ssec:apptoqs}}

In this section we will apply the results found in Secs. \ref{sec:pertmeth} and \ref{sec:gltd} to some charged spacetimes to demonstrate the effectiveness of the perturbative method, and to study in more  detail the effect of the electromagnetic interaction on the GEL observables. 

\subsection{The RN-type spacetimes \label{subsec:rnres}}

By RN-type spacetimes, here we include not only the well known RN BH spacetime but also any SSS spacetimes that have an exterior junction with the RN BH spacetime, including charged fluid \cite{Anninos:2001yb, Ivanov:2002jy}, thin-shell wormhole  \cite{Eiroa:2003wp}, gravastar \cite{Horvat:2008ch}, compact star \cite{Maurya:2015wma}, anisotropic compact object \cite{Morales:2018nmq,Sharif:2018toc} and polytropic sphere \cite{Arbanil:2013pua} solutions. Because the signals studied in this work only travel in the weak field limit, the gravity and electric interaction they experience in these spacetimes are exact the same as in RN BH spacetime carrying the same amount of charge.

The RN type spacetime
is the simplest charged spacetime, described by the following metric functions and electric potential
\be 
A(r)=\frac{1}{B(r)}=1-\frac{2M}{r}+\frac{Q^2}{r^2},~A_0(r)=-\frac{Q}{r}. \ee
Clear, they have particularly simple asymptotic expansion coefficients
\begin{subequations}
\label{eq:rncoeff}
\begin{align}
&a_0=1,~a_1=-2M,~a_2=Q^2,~a_n=0~~~(n\geq 3),\\
&b_0=1,~b_1=2M,~b_2=4M^2-Q^2,~\cdots,\\
&c_0=1,~c_n=0~~~(n\geq 1),\\
&a_{01}=-Q,~a_{0n}=0~~~(n\geq 1).
\end{align}
\end{subequations} 
Substituting these coefficients into \eqref{eq:ynexp} and then into \eqref{eq:dphifinal}, one finds the change of the angular coordinate in RN spacetime to the third order (second non-trivial order) 
\bea 
\Delta\phi_{\mathrm{R}} &=&
\sum_{i=s,d} \lcb
    \frac{\pi}2 -\theta_i + c_i \lb 1+\frac{1}{v^2} -\frac{\hat{q} \hat{Q} \sqrt{1-v^2}}{v^2}\rb  \frac{M}b\right.\nn\\
    &&+\frac18\lb \pi +2s_ic_i-2\theta_i\rb\lsb  3\lb 1+\frac{4}{v^2}\rb\right. \nn\\
    &&-\lb 1+\frac{2}{v^2}\rb\hat{Q}^2-\frac{12 \hat{q} \hat{Q} \sqrt{1-v^2}}{v^2}\nn\\
    &&\left.\left.+2\lb\frac{1}{v^2}-1\rb\hat{q}^2 \hat{Q}^2\rsb \frac{M^2}{b^2}+\mathcal{O}\lb \frac{M^3}{b^3}\rb \rcb, \label{eq:rnfirstfew}
\eea
where $\hat{Q}\equiv Q/M$ and the terms proportional to $\hat{q}^n~(n\geq 1)$ are the electromagnetic contributions. Setting $\hat{q}$ to zero, we recover the pure gravitational deflection in the RN spacetime (see Eq. (5.3) of Ref. \cite{Huang:2020trl}).
In the large $r_{s,d}$ limit, $\theta_i$ can be expanded and then $\Delta\phi$ becomes
\begin{align}
 \Delta\phi_{\mathrm{R}}=& \sum_{i=s,d}\lcb\frac{\pi}{2}+\lb 1+\frac{1}{v^2}-\frac{\hat{q} \hat{Q} \sqrt{1-v^2}}{v^2}\rb \frac{M}{b}\right.\nn\\
  &+\frac{\pi}{8}\lsb 3\lb 1+\frac{4}{v^2}\rb -\lb 1+\frac{2}{v^2}\rb \hat{Q}^2\right.\nn\\
&\left.-\frac{12 \hat{q} \hat{Q}\sqrt{1-v^2} }{v^2}
   +2\lb \frac{1}{v^2}-1\rb \hat{q}^2\hat{Q}^2\rsb\frac{M^2}{b^2}\nn\\
&\left.-\frac{b}{r_i}+\frac12\lb  \frac{1}{v^2}-1-\frac{\hat{q} \hat{Q} \sqrt{1-v^2}}{ v^2}\rb \frac{b^2 }{ r_i^2}\frac{M}{b}\rcb\nn\\
& +\mathcal{O}\lb \frac{M^3}{b^3},\frac{b^3}{r_i^3}\rb
\label{eq:rnfinite}
\end{align} 
The finite distance effect in this formula is more transparent than in Eq. \eqref{eq:rnfirstfew}. It is seen that at the leading order of $b/r_{s,d}$, the finite distance of the source/detector would slightly decrease the deflection angle, as expected from the fact that the gravitational bending of the trajectories is always towards the lens. 
In the infinite distance limit, this is further simplified to 
\begin{align}
 \Delta\phi_{\mathrm{R}}=&\pi+2\lb 1+\frac{1}{v^2}-\frac{\hat{q} \hat{Q} \sqrt{1-v^2}}{v^2}\rb \frac{M}{b}\nn\\
  &+\frac{\pi}{4}\lsb 3\lb 1+\frac{4}{v^2}\rb -\lb 1+\frac{2}{v^2}\rb \hat{Q}^2\right.\nn\\
 &\left.-\frac{12 \hat{q} \hat{Q}\sqrt{1-v^2} }{v^2}
   +2\lb \frac{1}{v^2}-1\rb \hat{q}^2\hat{Q}^2\rsb\frac{M^2}{b^2}\nn\\
& +\mathcal{O}\lb \frac{M^3}{b^3}\rb
\label{eq:rninf}
\end{align}
which agrees with Refs. \cite{Crisnejo:2019xtp,Li:2020ozr}. It is seen from Eq. \eqref{eq:rninf} that the spacetime charge $Q$ affects the deflection gravitationally at the order $\mathcal{O}(M/b)^2$ but electrically at one order lower, i.e., at the order $\mathcal{O}(M/b)$. Also note that setting $M=0$,  Eq. \eqref{eq:rninf} reduces to the pure relativistic electromagnetic deflection by a charge $Q$ and agrees with known result in Ref. \cite{bk:padamanabhan}. In this sense, setting $M=0$ in Eq. \eqref{eq:rnfinite} yields a relativistic Coulomb scattering deflection angle with finite distance effect of the source and detector taken into account. 

The apparent angles of the images in the RN spacetime are obtained by substituting the coefficients  \eqref{eq:rncoeff} into Eq. \eqref{eq:thetapmres}. The result is
\be
\theta_{\pm\mathrm{R}} =b_{\pm\mathrm{R}} \lsb \frac{1}{r_d}
+\lb \frac{-1}{v^2}+\frac{\hat{Q}\hat{q}\sqrt{1-v^2}}{v^2}\rb \frac{M}{r_d^2}\rsb 
\label{eq:thetapmresrn}
\ee
where $b_{\pm\mathrm{R}}$ are still given by Eq. \eqref{eq:blrtophi} but $\eta$ now takes the form
\be
\eta_{\mathrm{R}}
=\frac{8  (r_d+r_s)M}{\beta_L^2 r_d r_s}\lb\frac{1}{ v^2}+1-\frac{\hat{Q} \hat{q} \sqrt{1-v^2}}{v^2}\rb .
\label{eq:etainbetalrn}
\ee
The magnification is still given by Eq. \eqref{eq:magnif} with $u_\beta=\beta/\theta_{\mathrm{E}}$ but now $\theta_\mathrm{E}$ is updated to 
\be 
\theta_\mathrm{E,R}=\sqrt{\frac{2Mr_s}{r_d(r_s+r_d)}\lb\frac{1}{ v^2}+1-\frac{\hat{Q} \hat{q} \sqrt{1-v^2}}{v^2}\rb}.
\label{eq:thetaeresrn}
\ee

\begin{figure}[htp!]
    \centering
    \includegraphics{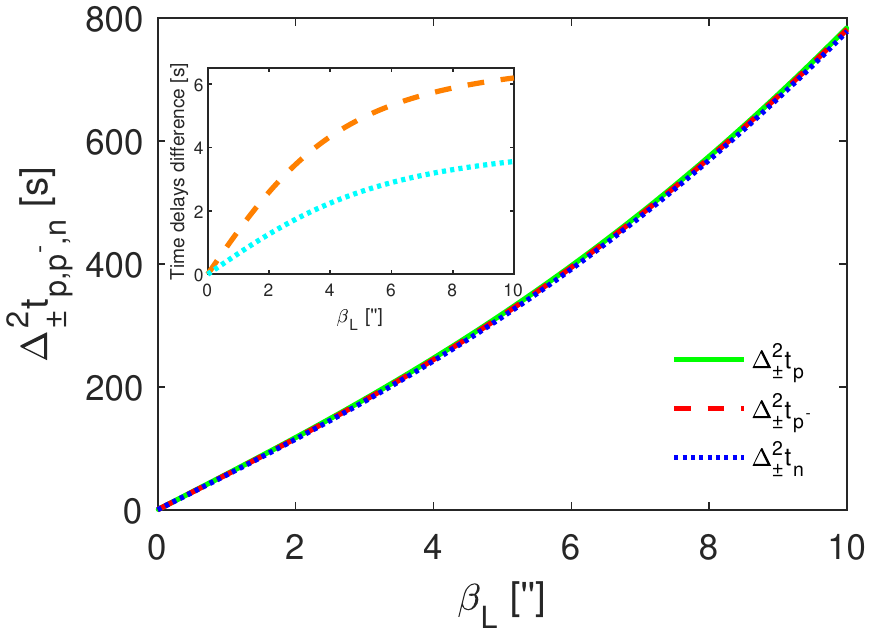}\\
    (a)\\
    \includegraphics{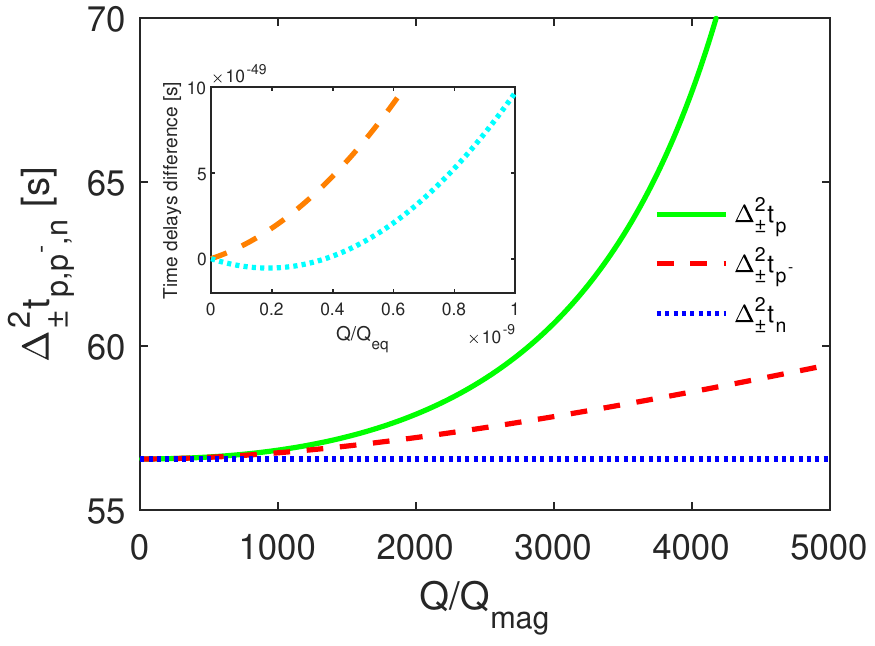}\\
    (b)
    \caption{The time delay $\Delta^2t_{\pm\mathrm{R}}$ using Eq. \eqref{eq:dtleadingrn}. (a) The dependence on $\beta_L$. All particles in this plot have the same velocity determined by $10^{19}$ [eV] protons. Inset: the differences between the proton time delay (dashed) or antiproton time delay (dotted) against the neutral particles.$Q$ was set to $2\times 10^3Q_{\mathrm{mag}}$ in this subplot.  (b) The dependence on $Q$ for $\beta_L=1$ $[^{\prime\prime}]$. Inset: dependence on $Q$ when $\beta$ is small ($\beta_L=10^{-10}$ $[^{\prime\prime}]$). }
    \label{fig:rnplottimedelay}
\end{figure}

To compute the time delay in the RN spacetime, we can substitute the coefficients \eqref{eq:rncoeff} into Eq. \eqref{eq:dtleading} and find 
\bea 
\Delta^2t_{\pm\mathrm{R}}
&=&
 \lb 1+v^2 - \hat{q}\hat{Q} \sqrt{1-v^2}\rb \frac{4M\sqrt{1+\eta}}{ v^3 \eta}\nn\\
&&+
2M\lsb \frac{1}{v^3}-\frac{3}{v} - \hat{q}\hat{Q} \left(\frac{1}{v^2}-1\right)^{\frac{3}{2}}\rsb\nn\\ &&\times  
 \ln \frac{\sqrt{1+\eta}-1}{\sqrt{1+\eta}+1}
+\mathcal{O}\lb \beta M,\frac{b^3}{r_{s,d}^2},\frac{M^2}{b}\rb 
\label{eq:dtleadingrn}
\eea
with $\eta$ given in Eq. \eqref{eq:etainbetalrn} or in terms of $\beta$ 
\be
\eta_{\mathrm{R}}(\beta)
=\frac{8  r_s M}{\beta^2 (r_d+r_s)r_d }\lb\frac{1}{ v^2}+1-\frac{\hat{Q} \hat{q} \sqrt{1-v^2}}{v^2}\rb .
\label{eq:etainbetarn}
\ee
From the illustration in Fig. \ref{fig:gel2}, we were clear that the flight times of charges with $\mathrm{sign}(qQ)=-1$ (or $=1$) from both sides will be larger (or smaller) than neutral signals, but the relative sizes of the time delays of the charged and neutral signal were not very transparent to us. In order to see how $\Delta^2 t_{\pm\mathrm{R}}$ was affected by the electromagnetic interaction, one can expand Eq. \eqref{eq:dtleadingrn} for small $\hat{q}\hat{Q}$ (note that $\eta$ depends on $\hat{q}\hat{Q}$ too). Carrying out this expansion, the time delay at small $\hat{q}Q$ and small $\beta$ becomes 
\bea 
&&\Delta^2 t_{\pm\mathrm{R}}(\beta\to0,\hat{q}\hat{Q}\to0)=\beta \sqrt{
\frac{2r_d(r_s+r_d)}{Mr_s}
}\lsb \frac{4M}{\sqrt{1+v^2}}\right.\nn\\
&&~~~~\left.+\hat{q}Q\lb \frac{1-v^2}{1+v^2}\rb^{3/2}\rsb +\mathcal{O}\lsb \beta^2,(\hat{q}\hat{Q})^2\rsb.\label{eq:tdsmallbeta}
\eea
It is seen that at the leading order, for $\mathrm{sign}(qQ)=1$ (or $\mathrm{sign}(qQ)=-1$),  the time delay is actually 
larger (or smaller) comparing to neutral particles. 
For signals with larger $\beta$ or $|qQ|$ however, numerical study shows that time delay might depend on $qQ$ differently (see Fig. \ref{fig:rnplottimedelay} (b)). 

\begin{figure}[htp!]
    \centering
    \includegraphics{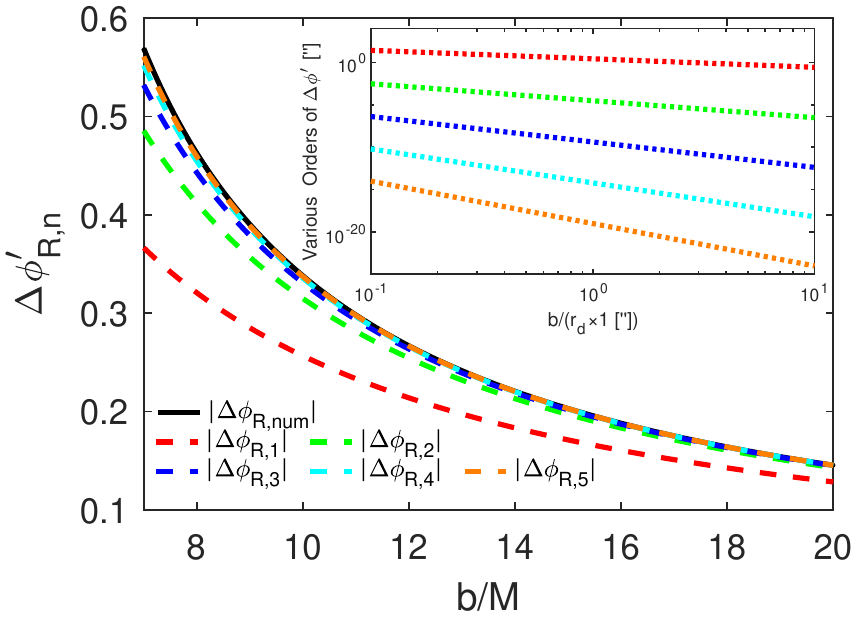}\\
    (a)\\
    \includegraphics{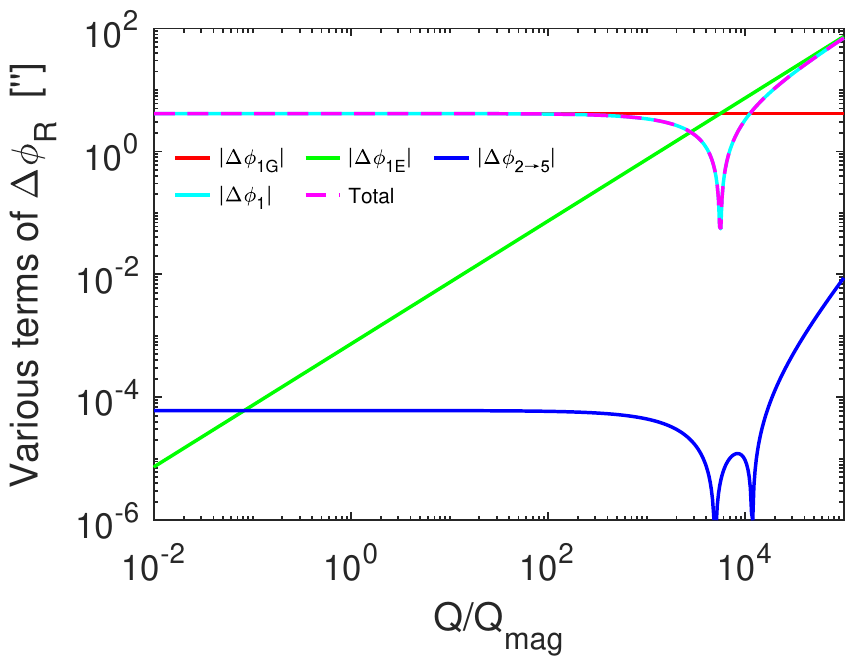}\\
    (b)
    \caption{The deflection angle $\Delta\phi^\prime_\mathrm{R}$ using Eq. \eqref{eq:rnfirstfew}. (a) Dash lines from bottom to top corresponds respectively to $\Delta \phi^\prime_\mathrm{R}$ truncated to $\mathcal{O}(M/b)^1$ to $\mathcal{O}(M/b)^5$ (note for clarity reason the 3rd-5th orders were not shown in Eq. \eqref{eq:rnfirstfew} but the third order is given in Eq. \eqref{eq:y3inrn}). The solid line is the numerical integration result. Inset: from top to bottom the contributions from order $\mathcal{O}(M/b)^1$ to $\mathcal{O}(M/b)^5$ in Eqs. \eqref{eq:rnfirstfew} for large $b$. $Q$ is set to $2\times 10^3Q_\mathrm{mag}$ in this subplot. (b) Various terms and orders of Eq. 
    \eqref{eq:rnfirstfew} as functions of $Q$. See text for details. $b$ is set to $r_d\cdot 1~ [^{\prime\prime}]$ in this subplot.}
    \label{fig:rnplots}
\end{figure}

\begin{figure}[htp!]
    \centering
    \includegraphics{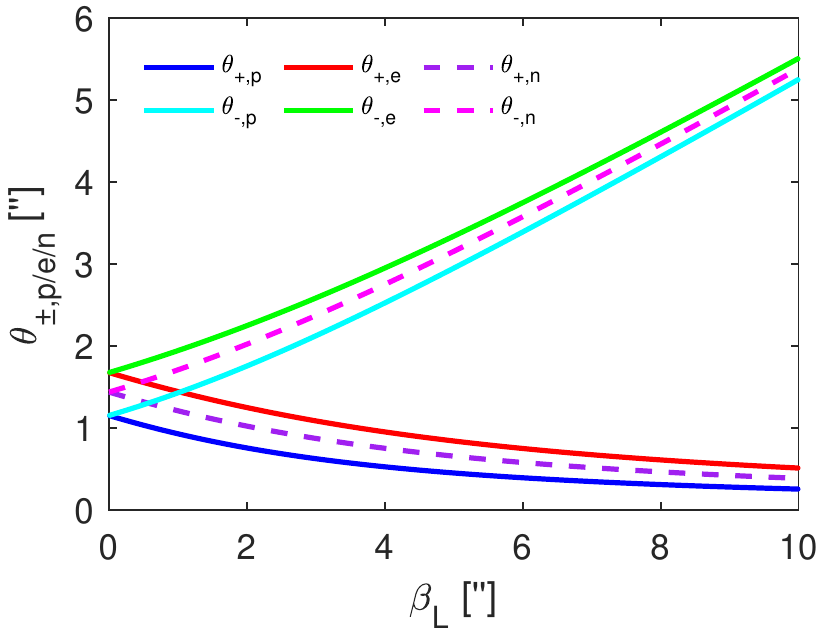}\\
    (a)\\
    \includegraphics{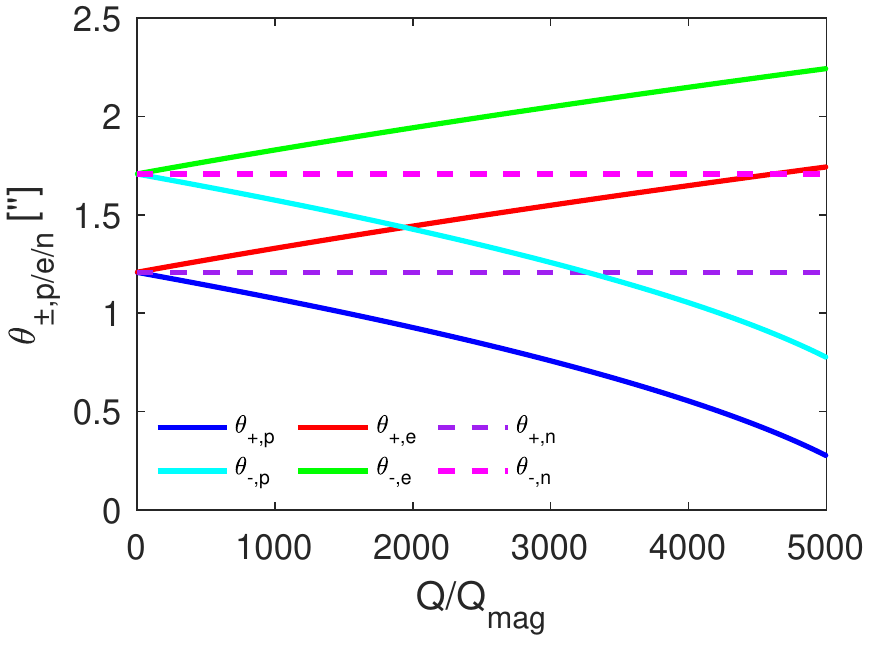}\\
    (b)
    \caption{The apparent angles $\theta_{\pm\mathrm{R}}$ using Eq. \eqref{eq:thetapmresrn}. (a) The apparent angles' dependence on $\beta_L$ for protons. For comparison, electrons with same energy and neutral particles with same velocity are also plotted. $Q$ is set to $2.0\times10^3 Q_{\mathrm{mag}}$ in this subplot. (b) The apparent angle as a function of $Q$. $\beta_L$ is set to 1 [$^{\prime\prime}$] in this subplot.}
    \label{fig:rnplotthetapm}
\end{figure}

\begin{figure}[htp!]
    \centering
    \includegraphics{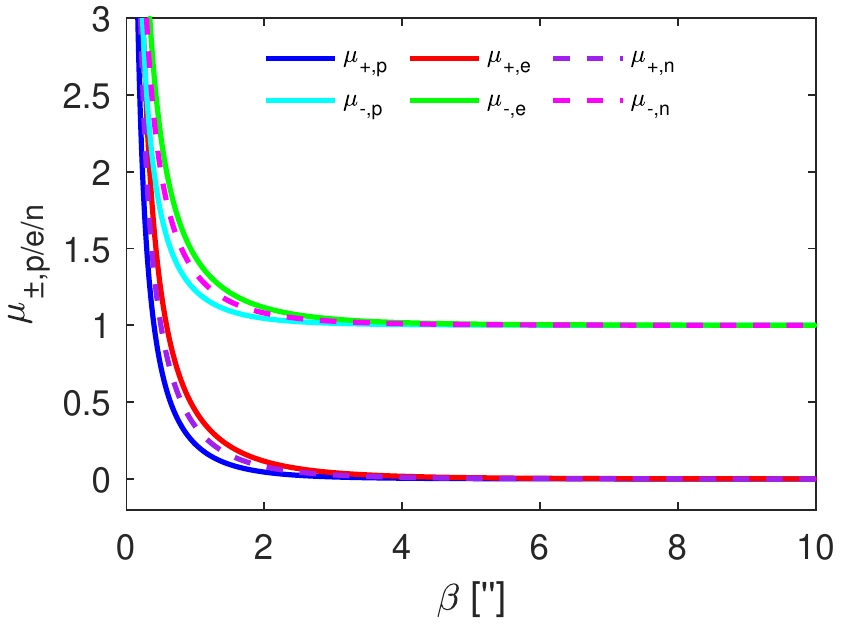}\\
    (a)\\
    \includegraphics{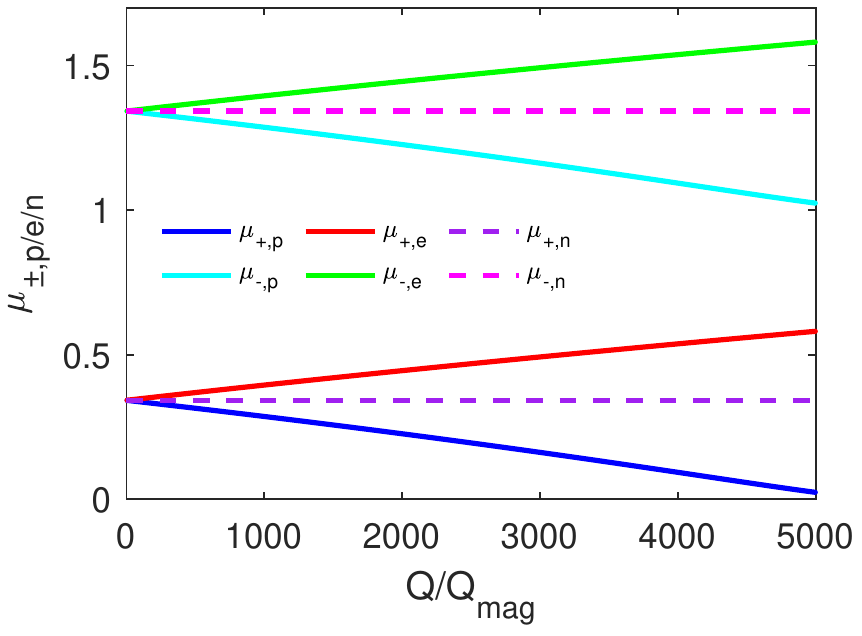}\\
    (b)
    \caption{The magnifications $\mu_{\pm\mathrm{R}}$ using Eq. \eqref{eq:magnif}. (a) Dependence of the magnification on $\beta$ for protons. For comparison, electrons with same energy and neutral particles with same velocity are also plotted. $Q$ is set to $2.0\times10^3Q_{\mathrm{mag}}$. (b) The magnification as a function of $Q$. $\beta$ is set to 1 [$^{\prime\prime}$] in this subplot. }
    \label{fig:rnmagplot}
\end{figure}

To verify the correctness of the above results and illustrate more clearly the effect of the electromagnetic interaction, in below we will plot these variables as functions of $b$, $\beta$ (or $\beta_L$) and $Q$. We assume that the deflection and GEL are caused by the Sgr A* SMBH, which carries a small charge. Although observationally the charge of the BH is not well constrained, theoretically there are at least three scales that can be used for reference. The first is related to the quasi-equilibrium of electrons and protons in the stellar atmosphere, which results in a small positive charge $Q_{\mathrm{eq}}$ of the astrophysical object \cite{eddington,bally}
\be 
Q_{\mathrm{eq}}\approx 100 \frac{M}{M_\odot}~[\text{C}]. 
\ee
If the charge of the BH is induced by the magnetic field $B_{\mathrm{mag}}$ around it, then its value should be around 
\begin{align}
Q_{\mathrm{mag}}
\approx 1.46\times 10^2\left(\frac{M}{M_\odot}\right)^2\frac{B_\mathrm{mag}}{10~[\text{G}]}~[\text{C}],
\end{align}
where $B_{\mathrm{mag}}$ is typically of order 10 [G]. Note that for typical SMBHs having mass $M\gtrsim \mathcal{O}(10^6M_\odot)$ both these two charges are much smaller than the third scale of $Q$, i.e., the extreme RN spacetime limit $Q_{\mathrm{extr}}$
\begin{align}
   Q_{\mathrm{eq}}\ll Q_{\mathrm{mag}}\ll  Q_{\mathrm{extr}}\approx 1.72\times 10^{20} \frac{M}{M_\odot}~[\text{C}].
\end{align}
Observing all these, and to see the electromagnetic effect more clearly and adequately, in this work we will limit the charge $Q$ to $\sqrt{Q_{\mathrm{mag}}Q_{\mathrm{extr}}}\approx 1.85\times 10^6Q_{\mathrm{mag}}\approx Q_{\mathrm{extr}}/(1.85\times 10^6)$.
Using the Sgr A* SMBH mass of $4.1\times 10^6 M_\odot$, distance  $r_d=8.1$ [kpc] and assuming $r_s=r_d$, and for proton energy of $10^{19}$ [eV] (which then fixes its velocity), we plotted the deflection angle $\Delta\phi^\prime_{\mathrm{R}}$ 
\be 
\Delta\phi^\prime_{\mathrm{R}}=
\Delta\phi_{\mathrm{R}}-\pi-\theta_s-\theta_d \label{eq:defangdef}
\ee
as functions of $b$ and $Q$ in Fig. \ref{fig:rnplots}, apparent angles $\theta_{\pm\mathrm{R}}$, magnifications $\mu_{\pm\mathrm{R}}$ and time delay $\Delta^2 t_{\pm\mathrm{R}}$ as functions of $\beta$ (or $\beta_L$) and $Q$ in Figs. \ref{fig:rnplotthetapm}, \ref{fig:rnmagplot} and \ref{fig:rnplottimedelay} respectively.

In Fig. \ref{fig:rnplots} (a), the deflection angle \eqref{eq:defangdef} with $\Delta\phi_{\mathrm{R}}$ from Eq. \eqref{eq:rnfirstfew} truncated to different orders of $M/b$, and the corresponding value obtained by direct numerical integration of Eq. \eqref{eq:deltaphi} are plotted as functions of $b$. The truncation order is labeled by subscript $n$. It is seen that as the truncation order increases, the accuracy of the series result also increases. For very large $b$, from the inset we see that the  contributions from each order also decreases rapidly as the order increases, as expected from the expansions of $\Delta\phi_{\mathrm{R}}$ given in Eqs. \eqref{eq:rnfinite} or \eqref{eq:rninf}. Note that since in these plots the $b/r_{s,d}$ are all very small, the finite distance effect of the source/distance can not be distinguished in them. In order to see the relative contribution of the electromagnetic interaction to $\Delta\phi^\prime_\mathrm{R}$, in Fig. \ref{fig:rnplots} (b) we plot as functions of $Q$ the leading order contribution from the gravitational interaction $\Delta\phi_{1G}$ and electromagnetic interaction $\Delta\phi_{1E}$, which correspond respectively to the $\lb 1+\frac{1}{v^2}\rb \frac{M}{b}$ term and $-\frac{\hat{q} \hat{Q} \sqrt{1-v^2}}{v^2}  \frac{M}b$ term in Eqs. \eqref{eq:rnfirstfew}, \eqref{eq:rnfinite} or \eqref{eq:rninf} (again, for the given $b$ and $r_{s,d}$, the finite distance effect can not be distinguished in the plot). The entire first order, $\Delta\phi_1$, the second to fifth order contribution, $\Delta\phi_{2\to 5}$, and the total $\Delta\phi^\prime_\mathrm{R}$ are also plotted. It is seen that the leading gravitational contribution dominates $\Delta\phi^\prime_\mathrm{R}$ when $Q\lesssim 5.58\times 10^3Q_\mathrm{mag}$. This value is indeed determined by the comparison between $\Delta\phi_{1G}$ and $\Delta\phi_{1E}$  of Eq. \eqref{eq:rnfirstfew}, i.e., $\mathcal{O}\lb \frac{qQ}{E}\rb \approx
\mathcal{O}\lb M\rb$, from which we obtain 
\be
 Q\approx \frac{ME}{q}\approx 10^3 Q_{\mathrm{mag}} \label{eq:qqdom}
\ee
for the chosen $M$ and signal energy $E$. 
If $Q$ grows larger than this value, the electromagnetic contribution will be larger than the gravitational term, until $Q$ reaches the value restricted by Eq. \eqref{eq:pertvalidcond} at which point the perturbative method breaks down. 
Note that since in this figure the electromagnetic interaction between the proton and the central charge ($Q>0$) is repulsive, when $Q$ exceeds the value given by Eq. \eqref{eq:qqdom}, the deflection angle $\Delta\phi^\prime_\mathrm{R}$ will changes sign. That is, the traditional converging GEL effect will not happen and signals diverge away from the center, similar to lensing using concave lenses. In contrast, if the $\mathrm{sign}(qQ)<0$, the converging lensing always happens, regardless the size of $Q$.

In Fig. \ref{fig:rnplotthetapm}, we plot the apparent angles $\theta_{\pm \mathrm{R}}$ in Eq. \eqref{eq:thetapmresrn} as functions of the source location $\beta_L$ and lens charge $Q$. From Fig. \ref{fig:rnplotthetapm} (a) we see that as $\beta_L$ increases from 0 to about 10 [$^{\prime\prime}$], the $\theta_{-\mathrm{R}}$ (or $\theta_{+\mathrm{R}}$) corresponding to the top (or bottom) trajectory in Fig. \ref{fig:glcharge} increases (or decreases). For comparison, we also plotted for electrons and neutral particles with the same energy. For these three kinds of particles, the values of $\theta_{\pm \mathrm{R}}$ at $\beta_L=0$  are the corresponding Einstein ring size at the fixed $Q$, as specified by Eq. \eqref{eq:thetaeresrn}. More importantly, it is seen that the apparent angles of electrons from both the top and bottom trajectories from any fixed $\beta_L$ are larger than those of neutral particles which in turn are larger than those of protons. This is in accord with the previous observation made after Eq. \eqref{eq:thetaeres} that if $\mathrm{sign}(qQ)=1$ (or $-1$), the $\theta_{\pm}$ will be smaller (or larger) than neutral particle. This is a fundamental effect of the electrostatic interaction on the GEL image apparent angles, which is also reflected in  Fig. \ref{fig:gel2}. Fig. \ref{fig:rnplotthetapm} (b) shows how these apparent angles evolve with the increase of $Q$. From this,  it is clear that quantitatively as $Q$ increases, the deviations of $\theta_{\pm\mathrm{R}}$ of charged signals from that of neutral particle also increase.

In Fig. \ref{fig:rnmagplot}, the magnifications of the images as given in Eq. \eqref{eq:magnif} with $\theta_{\mathrm{E,R}}$ in Eq. \eqref{eq:thetaeresrn} are plotted. It is seen from Fig. \ref{fig:rnmagplot} (a) that both $\mu_{\pm \mathrm{R}}$ decrease monotonically as $\beta$ departs from zero. This is indeed expected from Eq. \eqref{eq:magnif}
because a larger $\beta$ leads to larger $u_\beta$ which decreases $\mu_{\pm \mathrm{R}}$. On the other hand, it is seen from Fig. \ref{fig:rnmagplot} (b) that as predicted by the analysis after Eq. \eqref{eq:magnif}, the larger $Q$ leads to the smaller (or larger) $\theta_\mathrm{E,R}$ and consequently smaller (or larger) magnifications for protons (or electrons) images. 

Finally in Fig. \ref{fig:rnplottimedelay}, the time delay between the two images is plotted as a function of $\beta_L$ and $Q$ for UHE protons. To see the effect of particle charge (and only charge), we also plot the time delay of neutral particles and antiprotons. We choose to plot antiprotons rather than electrons because the time delay is sensitive to both velocity and energy of the signal, and only particles with the same mass can have both velocity and energy equal.
It is seen from Fig. \ref{fig:rnplottimedelay} (a) that for small $\beta_L$, the time delays for both the protons and antiprotons increase almost linearly as $\beta_L$ increases. This can be understood from the case \eqref{eq:dtetalarge} and that $\eta\sim \beta_L^{-2}$ from the relation \eqref{eq:etainbetal}. As  $\beta_L$ grows larger to about $\sim6$ [$^{\prime\prime}$], the condition \eqref{eq:largeetacond} starts to break down and the time delay \eqref{eq:dtleadingrn} deviates from Eq. \eqref{eq:dtetalarge}. A simple numerical evaluation shows that for the given range of $\beta_L$ (less than 10 [$^{\prime\prime}$]), the opposite limit $\eta\ll 1$ still can not be reached and one should still use Eq. \eqref{eq:dtleadingrn} to accurately describe the time delay for $\beta_L$ from $\sim 6$ to 10 [$^{\prime\prime}$]. To distinguish the time delays for these signals, in the inset we plotted the difference between time delays of charged signals and neutral particles. It is seen that for the chosen $Q=2\times 10^3Q_{\mathrm{mag}}$ and the chosen range of $\beta_L$, signals with $\mathrm{sign}(qQ)=1$ have a time delay larger than $\mathrm{sign}(qQ)=-1$  which in turn is larger than neutral particles time delay. For a large portion of the plotted $\beta_L$, their differences are both roughly a few seconds. This is much larger than typical observatory time resolutions such as neutrino or Gamma ray observatories, therefore allowing potential astrophysical applications, should the time delays of multimessenger events be observed. In addition, we note that the above ordering of the time delays is not a conflict with Eq. \eqref{eq:tdsmallbeta} because here the small $|qQ|$ condition is not satisfied. 

In Fig. \ref{fig:rnplottimedelay} (b), the effect of lens charge to the time delay is plotted. The inset shows the case that both $\beta_L$ and $|qQ|$ are small. It is seen that in this case, as predicted by Eq. \eqref{eq:tdsmallbeta}, the time delay for neutral particles is smaller than that of protons but larger than that of antiprotons with same velocity. When $Q$ becomes larger than $4.0\times 10^{-10}Q_{\mathrm{eq}}$ however, both the protons and antiprotons have larger time delay than neutral particles. This is also the case for even larger $Q$ and $\beta_L$, as shown in the main Fig. \ref{fig:rnplottimedelay} (b). 
In this range of $\beta_L$ and $Q$, the protons have larger time delay than antiprotons, and the time delay of neutral signals is the smallest. 

Having gained a basic understanding of the relevant physics of the above quantities, we can now check how these observables can be used if they were measured. Among all parameters involved, for the signal particles, usually their charge \mq and energy \mce can be completely determined at the observatory. For the spacetime, we will assume that the lens mass \mcm and its distance to the detector $r_d$ can be fixed by other astronomical means. For the GEL observables, usually the apparent angles $\theta_{+\mathrm{R}}$ and $\theta_{-\mathrm{R}}$, as well as the ratio between the two magnification $r_\mu\equiv \mu_{-\mathrm{R}}/\mu_{+\mathrm{R}}$ can be measured. The time delay $\Delta^2t_{\pm\mathrm{R}}$ sometimes can also be obtained, depending on the observation conditions of the GEL system \cite{gldatabase}. 
These observables indeed can be used to solve/constrain the lens/source parameters, such as the lens charge $Q$, source position $\beta$ (or $\beta_L$) and source distance $r_s$, which are often not directly measurable.

Using Eq. \eqref{eq:thetapmresrn} for the apparent angles of the image, Eq. \eqref{eq:dtleadingrn} for the time delay, and the corresponding values of $M, ~r_d$ for Sgr A* SMBH and $q,~E$ for UHE proton, we are indeed able to solve $Q, ~\beta$ and $r_s$ simultaneously. In Table \ref{tb:solvingqbrs}, we listed some typical values of $\theta_{\pm\mathrm{R}}$ and $\Delta^2 t_{\pm\mathrm{R}}$ and the solved values of $\beta,~Q$ and $r_s$. Unfortunately, due to the complexity of the above-mentioned equations, the solution  can only be expressed numerically.  
\begin{table*}[htp!]
    \centering
    \begin{tabular}{p{20mm}p{20mm}p{20mm}||p{20mm}p{20mm}p{20mm}}
\hline\hline
\multicolumn{2}{l}{Observables} &&\multicolumn{2}{l}{Solved quantities}\\
\hline
$\theta_{+\mathrm{R}}$ [$^{\prime\prime}$]&$\theta_{-\mathrm{R}}$ [$^{\prime\prime}$]  &$\Delta^2 t_{\pm,\mathrm{R}}~[M]$  &$\beta$ [$^{\prime\prime}$]&$Q~[10^{-9}M]$&$r_s$ [$r_d$]\\
\hline
 1.000 & 2.000 & 5.000 & 1.000 & 4.938 & 1.886 \\
 1.000 & 2.000 & 5.100 & 1.000 & 4.299 & 1.670 \\
 1.000 & 2.100 & 5.000 & 1.100 & 7.176 & 4.380 \\
 1.000 & 2.100 & 5.100 & 1.100 & 6.586 & 3.463 \\
 1.050 & 2.000 & 5.000 & 0.950 & 2.342 & 1.383 \\
 1.050 & 2.000 & 5.100 & 0.950 & 1.647 & 1.259 \\
 1.050 & 2.100 & 5.000 & 1.050 & 4.938 & 2.577 \\
 1.050 & 2.100 & 5.100 & 1.050 & 4.299 & 2.221 \\
\hline\hline
    \end{tabular}
    \caption{Constraining spacetime parameters and GEL configuration parameters using the observables. $M=4.1\times 10^6 M_\odot$ or $2.02\times 10^1$ [second] after conversion using natural units, and $r_d=8.1$ [kpc] for Sgr A* SMBH and $E=10^{19}$ [eV] for proton are used. Unit of $r_s$ is $r_d$. }
    \label{tb:solvingqbrs}
\end{table*}

\subsection{GMGHS spacetime results}

The GMGHS spacetime due to a charged dilaton is famous as a stringy BH spacetime. Deflection and lensing of neutral lightray in the weak field limit in this spacetime have been considered multiple times  \cite{Bhadra:2003zs,Keeton:2005jd,Ovgun:2018prw}. Recently, motion of charged test particles is also studied in this spacetime, concentrating on its inner-most stable circular orbit or BH shadow \cite{Ong:2019glf}. Here we consider the deflection and weak lensing of charged test particles in this spacetime for the first time (to our best knowledge).

The full GMGHS metric functions and four-potential are given by \cite{Garfinkle:1990qj,Bhadra:2003zs}
\bea
&&A(r) =\frac{1}{B(r)}=1 - \frac{2 M} {r},~C(r)=r\left( r-\frac{Q^2 \mathrm{e}^{-2\phi_0}}{M}\right),\nn\\
&&A_0(r)=-\frac{Q}{r},
\eea 
where $\phi_0\geq 0$ is the asymptotic value of the dilaton field. 
Apparently, the asymptotic coefficients are
\begin{subequations}
\label{eq:ghgmsexp}
	\begin{align}
		&a_{0}=1, ~a_{1}=-2M, ~a_n=0~~~ (n\geq 2),\\
		&b_{0}=1, ~b_{1}=2M, ~b_{2}=4 M^2, ~\cdots,\\
		&c_{0}=1, ~c_{1}=-\frac{Q^2 e^{-2 \phi _0}}{M}, ~c_n=0~~~(n\geq 2),\\
		&a_{01}=-Q, ~a_{0n}=0~~~(n\geq 2).
	\end{align}
\end{subequations}
Substituting these into Eq. \eqref{eq:ynexp} and then  \eqref{eq:dphifinal}, the change of the angular coordinate becomes
\begin{align}
		\Delta \phi_\mathrm{G} =&\sum_{i=s,d}\lcb \frac{\pi}{2}-\theta_i+c_i\left( 1+\frac{1}{v^2}-\frac{\hat{q} \hat{Q} \sqrt{1-v^2}}{v^2} \right) \frac{M}{b}\right.\nonumber \\
		&+\frac{1}{8} \left( \pi+2 s_i c_i-2\theta_i\right) \lsb 3\left(1+\frac{4}{v^2} \right)
		\right.\nonumber \\
		&-e^{-2 \phi_0}  \hat{Q}^2 \left(1+\frac{2}{v^2} \right)-\frac{e^{-4\phi_0}\hat{Q}^4 }{4}\nonumber  \\
		&-2\left(6-e^{-2 \phi_0 } \hat{Q}^2\right)\frac{\hat{Q} \hat{q} \sqrt{1-v^2}}{v^2} \nn\\
		&\left.\left.+2\left(\frac{1}{v^2}-1 \right) \hat{q}^2 \hat{Q}^2\rsb \frac{M^2}{b^2}+\mathcal{O}\lb\frac{M^3}{b^3}\rb\rcb \label{eq:dphigmghs}
	\end{align}
	
Comparing to the RN result, Eq. \eqref{eq:rnfirstfew}, we see from the $\mathcal{O}(M/b)^2$ order that the gravitational contribution of $Q$ to $\Delta\phi$ now is tuned by the asymptotic dilaton factor $e^{-2\phi_0}$, which is inherited from the coefficient $c_1$. In contrast, the electromagnetic contribution of $Q$ to $\Delta\phi$ appears only in one term in the second order of $\mathcal{O}(M/b)^2$, i.e., the $6-e^{-2\phi_0}\hat{Q}^2$ term. Since usually we have $\hat{Q}\leq 1$, the effect of $e^{-2\phi_0}$ on the electromagnetic interaction is much weaker than that on the gravitational deflection. Although some works used a zero $\phi_0$  \cite{Horowitz:1992ya,Blaga:2014spa}, indeed this is a variable not well constrained theoretically. When $\phi_0$ is large, then it is seen that the gravitational contribution from the charge to $\Delta\phi$ is turned off while the electromagnetic one is still present, unaltered at the leading order. 

Expanding Eq. \eqref{eq:dphigmghs} to the leading two orders of $b/r_{s,d}$, the finite distance $\Delta\phi$ that is parallel to Eq. \eqref{eq:rnfinite} in the GMGHS case becomes
\begin{align}
		\Delta \phi_\mathrm{G}=& \sum_{i=s,d} \lcb\frac{\pi}{2}+\left(1+\frac{1}{v^2}-\frac{\hat{q} \hat{Q} \sqrt{1-v^2}}{v^2} \right)\frac{M}{b}\right.\nonumber \\
		&+\frac{\pi}{8}\lsb3\left(1+\frac{4}{v^2} \right)-\left(1+\frac{2}{v^2} \right)e^{-2 \phi_0} \hat{Q^2}\right.\nonumber\\
		& -\frac{e^{-4\phi_0}  \hat{Q}^4}{4}-2\left(6- e^{-2 \phi_0}\hat{Q}^2  \right)\frac{ \hat{q} \hat{Q} \sqrt{1-v^2}}{v^2}\nonumber \\
		&\left.+2 \left(\frac{1}{v^2}-1 \right) \hat{q}^2  \hat{Q}^2\rsb \frac{M^2}{b^2}-\frac{b}{r_i}\nonumber\\
		&+\frac12 \left[\frac{1}{ v^2}-1-\frac{\hat{q} \hat{Q} \sqrt{1-v^2}}{v^2}-e^{-2 \phi_0}\hat{Q}^2  \right]\frac{b^2}{ r_i^2}\frac{M}{b}\nonumber\\
		&+\mathcal{O}\lb \frac{M^3}{b^3},\frac{b^3}{r_i^3}\rb.
	\end{align}
The infinite $r_{s,d}$ result is given by the first three terms 
	\begin{align}
		\Delta \phi_\mathrm{G} = &\pi+2\left(1+\frac{1}{v^2}-\frac{\hat{q} \hat{Q} \sqrt{1-v^2}}{v^2} \right)\frac{ M}{b}\nonumber \\
		&+\frac{ \pi}{4}\lsb 3\left(1+\frac{4}{v^2} \right)-\left(1+\frac{2}{v^2} \right) e^{-2 \phi_0} \hat{Q^2}\right.\nonumber\\
		& -\frac{ e^{-4\phi_0}  \hat{Q}^4}{4}-2\left(6-e^{-2 \phi_0}\hat{Q}^2  \right)\frac{ \hat{q} \hat{Q} \sqrt{1-v^2}}{2 v^2}\nonumber \\
		&\left.+2 \left(\frac{1}{v^2}-1 \right) \hat{q}^2  \hat{Q}^2\rsb \frac{M^2}{b^2}+\mathcal{O}\lb \frac{M^3}{b^3}\rb 
	\end{align}

For the apparent angles $\theta_{\pm\mathrm{G}}$ of the GELed images in this spacetime, again substituting the expansion coefficients \eqref{eq:ghgmsexp} into Eq.  \eqref{eq:thetapmres}, we obtain the result \be 
\theta_{\pm \mathrm{G}} =b_{\pm\mathrm{G}} \lsb \frac{1}{r_d}
+\lb \frac{-1}{v^2}+\frac{\hat{Q}e^{-2\phi_0}}{2}+\frac{\hat{Q}\hat{q}\sqrt{1-v^2}}{v^2}\rb \frac{M}{r_d^2}\rsb 
\label{eq:thetapmresgmghs}
\ee 
with the 
impact parameters 
$b_{\pm \mathrm{G}}$ still given by Eq. \eqref{eq:blrtophi} but with $ 
\eta_\mathrm{G}$ taking the same form as $\eta_\mathrm{R}$ in Eq. \eqref{eq:etainbetalrn} since both the RN and GMGHS spacetimes have the same $a_1,~b_1$ and $a_{01}$ which fix $\eta$ according to Eq. \eqref{eq:etainbetal}. For exactly the same reason, for the Einstein ring size $\theta_{\mathrm{E,G}}$, the magnifications $\mu_{\pm\mathrm{G}}$ and the time delay $\Delta^2t_{\pm\mathrm{G}}$, which should be obtained by substituting $a_1,~b_1$ and $a_{01}$ into Eqs. \eqref{eq:thetaeres}, \eqref{eq:magnif} and \eqref{eq:dtleading} respectively, their formulas in the GMGHS spacetime are the same as in the RN-type spacetimes. In other words, these three quantities in the GMGHS spacetime are given by Eqs. \eqref{eq:thetaeresrn}, \eqref{eq:magnif} with $u_\beta=\beta/\theta_{\mathrm{E,G}}$ and \eqref{eq:dtleadingrn} respectively.

\subsection{Charged Horndeski theory result}

Another spacetime that will be considered here is the charged Horndeski theory, whose metric functions and electric potential function are \cite{Cisterna:2014nua,Feng:2015wvb,Wang:2019cuf}
\bea 
&&A(r)=1 - \frac{2 M} {r}+\frac{Q^2}{4r^2}-\frac{Q^4}{192r^4},~B(r)=\frac{1}{A(r)}\lb 1-\frac{Q^2}{8r^2}\rb^2,\nn\\
&&C(r)=r^2,~A_0(r)=-\frac{Q}{r}+\frac{Q^3}{24r^3}.
\eea 
Only when $Q^2<9M^2/2$, this describes a BH spacetime. The above have the following asymptotic expansions 
\begin{subequations}
\label{eq:hcoeff}
\begin{align}
		&a_{0}=1, ~a_{1}=-2M,~ a_{2}=\frac{Q^2}{4},~a_{3}=0,\nn\\
		&~~~~~~~~~~~a_{4}=\frac{-Q^2}{192},~~~a_n=0~(n\geq5), \\
		&b_{0}=1,~ b_{1}=2M, ~b_{2}=4M^2 - \frac{Q^2}{2},~\cdots,\\
		&c_{0}=1, ~c_{n}=0~~~(n\geq1),\\
		&a_{01}=-Q,~ a_{02}=0, ~a_{03}=\frac{Q^3}{24},~a_{0n}=0~~~(n\geq4).   
	\end{align}
\end{subequations}
Up to the second order, this set of asymptotic coefficients are very similar to the RN coefficients \eqref{eq:rncoeff}. The only differences, which appear in $a_2$ and $b_2$ ($a_2$ here is scaled by a factor of $1/4$ and the $Q^2$ in $b_2$ is scaled by $1/2$), are quantitative but not qualitative.

Substituting these coefficients into  Eq. \eqref{eq:ynexp} and then  \eqref{eq:dphifinal}, we obtain the change of the angular coordinate
\begin{align}
		\Delta \phi_\mathrm{H} =&\sum_{i=s,d}\lcb \frac{\pi}{2}-\theta_i+c_i \left(1+\frac{1}{v^2}-\frac{\hat{q} \hat{Q} \sqrt{1-v^2}}{v^2}\right)\frac{M}{b}\right. \nonumber\\
		&+\frac{1}{8} \left( \pi+2 s_i c_i-2\theta_i\right)\lsb 3\left(1+\frac{4}{v^2}\right)-\left(1+\frac{1}{v^2}\right)\frac{\hat{Q}^2}{2}\right.\nonumber \\
		&\left.-\frac{12  \hat{q} \hat{Q} \sqrt{1-v^2}}{v^2}+2\left(\frac{1}{v^2}-1\right) \hat{q}^2\hat{Q}^2\rsb \frac{M^2}{b^2} \nn\\
		&\left.+\mathcal{O}\lb\frac{M^3}{b^3}\rb\rcb.
\end{align}
Comparing to the RN case \eqref{eq:rnfirstfew}, the difference of this $\Delta\phi$ is in the second-order terms, caused by the numerical factor difference in $a_2$ and $b_2$.
Its small $b/r_{s,d}$ expansion to the second order is given by 	\begin{align}
		\Delta \phi_\mathrm{H} =&\sum_{i=s,d}\lcb 
		\frac{\pi}{2}+\left(1+\frac{1}{v^2}-\frac{\hat{q} \hat{Q} \sqrt{1-v^2}}{v^2}\right)\frac{M}{b}\right.\nn\\ 
		&+\frac{\pi}{8} \lsb 3\left(1+\frac{4}{v^2} \right)-\left(1+\frac{1}{v^2}\right)\frac{\hat{Q}^2}{2}\right.\nn\\
		&\left.-\frac{12  \hat{q} \hat{Q} \sqrt{1-v^2}}{v^2}+2\left(\frac{1}{v^2}-1\right) \hat{q}^2\hat{Q}^2  \rsb \frac{M^2}{b^2} \nn\\
		&-\frac{b}{r_i}+\frac12 \left(\frac{1}{ v^2}-1-\frac{\hat{q} \hat{Q}\sqrt{1-v^2}}{v^2} \right)\frac{b^2}{r_i^2} \frac{M}{b}\nn\\
		&\left.+\mathcal{O}\lb \frac{M^3}{b^3},\frac{b^3}{r_i^3}\rb\rcb,
	\end{align}
and the corresponding infinite distance result is simply
	\begin{align}
		\Delta \phi_\mathrm{H} =& 
		\pi+2\left(1+\frac{1}{v^2}-\frac{\hat{q} \hat{Q} \sqrt{1-v^2}}{v^2}\right)\frac{M}{b}\nn\\ 
		&+\frac{\pi}{4} \lsb 3\left(1+\frac{4}{v^2} \right)-\left(1+\frac{1}{v^2}\right)\frac{\hat{Q}^2}{2}\right.\nn\\
		&\left.-\frac{12 \hat{q} \hat{Q} \sqrt{1-v^2}}{v^2}+2\left(\frac{1}{v^2}-1\right)\hat{q}^2 \hat{Q}^2 \rsb \frac{M^2}{b^2} \nn\\
		&+\mathcal{O}\lb \frac{M^3}{b^3},\frac{b^3}{r_i^3}\rb.
	\end{align}
Setting $\hat{q}=0$ and taking $v=1$, this further reduces to the deflection of chargeless null rays, and it agrees with Eq. (3.1) of Ref. \cite{Wang:2019cuf}.

For GEL by the Horndeski black hole, the impact parameters $b_{\pm \mathrm{H}}$ and other observables including apparent angles $\theta_{\pm \mathrm{H}}$, magnifications $\mu_{\pm \mathrm{H}}$ and time delay $\Delta^2 t_{\pm\mathrm{H}}$, can be obtained respectively from Eqs. \eqref{eq:blrtophi}, \eqref{eq:thetapmres}, \eqref{eq:magnif}, \eqref{eq:dtleading}. One important thing to notice is that these equations are all given to the leading order(s) and therefore are determined by the expansion coefficients up to order $a_1,~b_1,~c_1$ and $a_{01}$ of the metric functions and four-potential. Since to this order, coefficients for the 
Horndeski BH as given in Eq. \eqref{eq:hcoeff} have the same form as in the RN spacetime given in Eq. \eqref{eq:rncoeff}, the above mentioned quantities in the Horndeski case would also take the same form as in the RN case. In other words, $b_{\pm \mathrm{H}}$  are still given by Eq. \eqref{eq:blrtophi} with \eqref{eq:etainbetalrn}, $\theta_{\pm \mathrm{H}}$ by Eq. \eqref{eq:thetapmresrn}, $\mu_{\pm \mathrm{H}}$ by Eq. \eqref{eq:magnif} with \eqref{eq:thetaeresrn} and finally $\Delta t^2_{\pm\mathrm{H}}$ by Eq. \eqref{eq:dtleadingrn}. If one is interested in the higher-order difference between these quantities in the Horndeski and RN spacetimes, higher orders have to be pursued in the general formulas for these quantities. Although this is technically possible, we will not continue along this direction because it seems not much more physics can be gained. 

\section{Conclusions and discussions}

In this work, we considered using a perturbative approach the deflection angle and dual lensing of  charged signals due to the gravitational and electromagnetic interactions in the weak field limit in general charged SSS spacetime. Our method can address not only the bending and GEL of null but also timelike particles, not only charged but also neutral particles. Moreover, the finite distance effect of the source/observer is also taken into account, which allows us to solve an exact lensing equation. 

It is found in Eqs. \eqref{eq:y1exp} and \eqref{eq:dphifinal3} that the change of the angular coordinate $\Delta\phi$ receives an electromagnetic modification from the first non-trivial order. Depending on the sign of $qQ$, this electromagnetic modification might increase (for $\mathrm{sign}(qQ)=-1$) or decrease (for $\mathrm{sign}(qQ)=1$) the deflection angle by the amount determined by the size of $qQ/E$. In GEL, the extra electromagnetic interaction will decrease (or increase) the impact parameters and apparent angles of both images and the Einstein ring size if $\mathrm{sign}(qQ)=1$ (or $\mathrm{sign}(qQ)=-1$), according to Eqs. 
\eqref{eq:blrtophi} to \eqref{eq:thetaeres}. 
This is intuitively understandable from the fact that the same (or different) signs of $q$ and $Q$ will cause an extra repulsion (or attraction) between the lens and the signal and therefore the impact parameter has to adjust oppositely in order for the signals from the same source to reach the same observer. 

For the total travel time, for the same reasoning as in the case of deflection angle, $\mathrm{sign}(qQ)=1$ (or $\mathrm{sign}(qQ)=-1$) will decrease (or increase) $\Delta t$ for signals from both sides. The more interesting quantity here is effect of electromagnetic interaction on the time delay between the two images, which is the difference between two decreased or increased total travel times. It is found that for extremely small $\beta$ and $|qQ|$, comparing to neutral particles, the positive (or negative) $\mathrm{sign}(qQ)$ actually will increase (or decrease) the time delay. For larger $qQ$ or $\beta$ however, time delay of both positive and negative charges will be larger than that of the neutral particle with same velocity. 

The application of the results to three charged spacetimes in Sec. \ref{ssec:apptoqs}, especially to the RN spacetime, confirms the above conclusions drawn according to the results in general SSS spacetime. The charge of the spacetime we numerically considered is much smaller than its mass, and the charged particles are mainly the UHE protons in cosmic rays, although sometimes for comparison purpose electrons and neutral particles are also considered. We also applied the results to the GHGMS and charged Horndeski spacetimes. 

\acknowledgements
X. Xu and T. Jiang thank Mr. Zonghai Li and Mr. Ruizhe Liang for valuable discussions. This work is supported by the NNSF China 11504276 and MOST China 2014GB109004.

\appendix

\section{Integral formulas \label{sec:appd}}

When integrating Eqs.  \eqref{eq:phiinu} and \eqref{eq:tinu} with Eq. \eqref{eq:ynexp}, and Eq. \eqref{eq:znexp}, we need to compute integrals of the following form 
\be 
I_n(\theta_i)=\int_{\sin \theta }^1\frac{u^n}{\sqrt{1-u^2}}\dd u
~~(n=-2,~-1,~\cdots). \ee
This integral can be carried out using a change of variables $u=\sin\xi$ to find an elementary expression
\bea
&&I_n(\theta_i)=\int_\theta^{\pi/2}\sin^n \xi\dd \xi\nonumber\\
&=&
\begin{cases}
\cot\theta_i, ~~~~~~~~ ~~~~~~~~~~ n=-2,\\
\ln \lsb \cot\lb \frac{\theta_i}{2}\rb \rsb, ~~~~~~~~~ n=-1,\\
\displaystyle  \frac{(n-1)!!}{n!!} \left(\frac{\pi}{2}-\theta_i
    +\cos\theta_i\sum_{j=1}^{[\frac{n}{2}]} \frac{(2j-2)!!} {(2j-1)!!}\sin^{2j-1} \theta_i\right),\\
~~~~~~~~~~~~~~~~~~~~~~~~~~n=0,~2,~\cdots,\\
\displaystyle  
\frac{(n-1)!!}{n!!}\cos\theta_i \left(1
    +\sum_{j=1}^{[\frac{n}{2}]} \frac{(2j-1)!!}{(2j)!!} \sin^{2j}\theta_i\right),\\
~~~~~~~~~~~~~~~~~~~~~~~~~~n=1,~3,~\cdots.
\end{cases}
\label{eq:inthetares}
\eea
In the $\theta=0$ limit, i.e., the infinite source/detector distance limit, $I_n$ for non-negative $n$ can be further simplified
\be
I_n(0)=
\frac{(n-1)!!}{n!!}\begin{cases}
\displaystyle  \frac{\pi}{2},&~n=0,~2,~\cdots,\\
\displaystyle  1,&~n=1,~3,~\cdots.
\end{cases} \label{eq:inthetareslim}
\ee

To present the effect of finite source/detector distance to the deflection angle and time delay in a more transparent way, we will need the expansion of the first few $I_n(\theta_i)$ in the limits of small $b/r_i$ and $M/b$. They can be obtained by expanding $\theta_i$ in Eq. \eqref{eq:appang} for any given $b$ and find
\begin{subequations}
\label{eq:ifirstthreeexp}
\begin{align}
I_{-2}=&\frac{r_i}{b}-\frac{b}{2 r_i} -\lb\frac{a_1}{2v^2}-\frac{c_1}{2}-\frac{a_{01}\hat{q}\sqrt{1-v^2}}{v^2}\rb \frac{1}{b}\nn\\
&+ \mathcal{O}\lb \frac{b^2}{r_i^2}, \frac{M^2}{b^2}\rb , \\
I_{-1}=& -\ln \lb\frac{b}{2 r_i}\rb - \lb\frac{a_1}{2v^2}-\frac{c_1}{2}-\frac{a_{01}\hat{q}\sqrt{1-v^2}}{v^2}\rb\frac{b}{r_i} \frac{1}{b} \nn\\
&+ \mathcal{O}\lb \frac{b^2}{r_i^2}, \frac{M^2}{b^2}\rb, \\
I_0=&\frac{\pi}{2}-  \frac{b}{r_i}- \lb\frac{a_1}{2v^2}-\frac{c_1}{2}-\frac{a_{01}\hat{q}\sqrt{1-v^2}}{v^2}\rb\frac{b}{r_i^2}\nn\\
&+\mathcal{O}\lb \frac{b^3}{r_i^3}, \frac{M^2}{b^2}\rb , \label{eq:asyln}\\
I_1=&1-\frac{b^2}{2r_i^2}+ \mathcal{O}\lb \frac{b^3}{r_i^3}, \frac{M^2}{b^2}\rb,\\
I_2=&\frac{\pi}{4}+ \mathcal{O}\lb \frac{b^3}{r_i^3}, \frac{M^2}{b^2}\rb.
\end{align}
\end{subequations}

\section{Integral formulas \label{sec:appdyzhigh}}

The coefficients $y_n$ and $z_n$ in Eqs. \eqref{eq:dphifinal} and \eqref{eq:dtfinal} essentially  determine the results for the change of angular coordinate $\Delta \phi$ and time coordinate $\Delta t$. Besides those given in Eqs. \eqref{eq:ynexp} and \eqref{eq:znexp}, here we list one more order of them. That is, 
\begin{align}
y_3= & \left(\frac{1}{16 v^6}-\frac{3}{4 v^4}-\frac{3}{2 v^2}\right)a_1^3\nn\\
&+ \left(\frac{3}{16 v^4}+\frac{3}{4 v^2}\right) a_1^2\left(b_1+2 c_1\right)+ \left[\frac{3 b_1 \left(b_1-4 c_1\right)}{16 v^2}\right.\nn\\
&\left.+ \left(\frac{1}{4 v^4}+\frac{1}{v^2}\right)3 a_2-\frac{3 \left(b_2+2 c_2\right)}{4 v^2}\right]a_1\nn\\
&+ \left[\frac{b_1 \left(b_1-2 c_1\right)}{16}+ \left(-\frac{3 a_2}{4v^2}-\frac{b_2}{4}+\frac{ c_2}{2}\right)\right] b_1\nn\\
&+ \left(-\frac{3 a_2}{2 v^2}+\frac{b_2}{2}\right)c_1-\frac{3 a_3}{2 v^2}+\frac{b_3}{2}+c_3\nn\\
&- a_{01} \hat{q}\sqrt{1-v^2} \lcb \left(\frac{3}{8v^6}-\frac{3}{v^4}-\frac{3}{v^2}\right) a_1^2 \right.\nn\\
&+\left(\frac{3}{4 v^4}+\frac{3}{2 v^2}\right) \lsb a_1 \left(b_1+2 c_1\right)+2 a_2\rsb \nn\\
&\left.+\frac{3}{8 v^2}  \lsb b_1 \left(b_1-4 c_1\right)-4 b_2-8 c_2\rsb\rcb \nn\\
&-3  \left[a_1\left(\frac{1}{2v^4}+\frac{1}{v^2}\right)-\frac{b_1+2 c_1}{2 v^2}\right]a_{02}  \hat{q} \sqrt{1-v^2}\nn\\
&+\frac{3  a_{03}  \hat{q} \sqrt{1-v^2}}{v^2}+3 a_{01}\hat{q}^2\left(1-v^2\right)\nn\\
&\times  \lcb a_1 a_{01} \left(\frac{1}{4 v^6}-\frac{5}{4 v^4}-\frac{1}{2v^2}\right)\right.\nn\\
&\left.+\left(\frac{1}{v^4}+\frac{1}{v^2}\right) \left(a_{01} \lsb b_1+2 c_1\right)+4a_{02}\rsb \rcb \nn\\
&- \left(\frac{1}{2 v^6}-\frac{3}{2 v^4}\right)a_{01}^3 \hat{q}^3 \left(1-v^2\right)^{3/2},
\label{eq:y3general}
\end{align}
and
\begin{align}
z_2=&-\left(\frac{1}{16 v^7}-\frac{3}{8 v^5}+\frac{3}{2 v^3}+\frac{1}{v}\right)a_1^3\nn\\
&- \left(\frac{1}{16 v^5}-\frac{1}{2 v^3}-\frac{1}{2 v}\right) a_1^2\left(b_1+2 c_1\right)\nn\\
&+ \left[ \left(-\frac{1}{4 v^5}+\frac{2}{v^3}+\frac{2}{v}\right)a_2\right.\nn\\
&+ \left(\frac{1}{16 v^3}+\frac{1}{8 v}\right)b_1\left(b_1-4c_1\right)\nn\\
&\left.-\left(\frac{1}{4 v^3}+\frac{1}{2 v}\right) \left(b_2+2 c_2\right)\right]a_1 +\frac{b_1^3}{16 v}-\frac{b_1^2 c_1}{8 v}\nn\\
&- \left[ \left(\frac{1}{4 v^3}+\frac{1}{2 v}\right)a_2+\frac{1}{4 v} \left(b_2-2 c_2\right)\right]b_1\nn\\
&+\left[ \left(-\frac{1}{2v^3}-\frac{1}{v}\right)a_2+\frac{b_2}{2 v}\right]c_1\nn\\
&- \left(\frac{1}{2v^3}+\frac{1}{v}\right)a_3+\frac{1}{2 v}\left(b_3+2 c_3\right)\nn\\
&+\frac{ a_{01}\hat{q}\sqrt{1-v^2}}{v} \lcb \left(\frac{3}{8v^6}-\frac{13}{8v^4}+\frac{4}{v^2}+1\right) a_1^2 \right.\nn\\
&+\left(\frac{1}{4 v^4}-\frac{5}{4 v^2}-\frac{1}{2}\right)  \lsb a_1 \left(b_1+2 c_1\right)+2 a_2\rsb \nn\\
&\left.- \left(\frac{1}{8 v^2}+\frac{1}{8}\right)  \left[b_1\left(b_1-4 c_1\right)-4 b_2-8 c_2\right]\rcb\nn\\
&+\left[ \left(\frac{1}{2v^4}-\frac{5}{2v^2}-1\right)a_1+\left(\frac{1}{v^2}+1\right) \left(b_1+2 c_1\right)\right]\nn\\
&\times \frac{a_{02} \hat{q} \sqrt{1-v^2} }{2 v}+\left(\frac{1}{v^2}+1\right) \frac{a_{03} \hat{q} \sqrt{1-v^2}}{v}\nn\\
&+a_{01} \hat{q}^2\left(1-v^2\right)\left[\left(-\frac{3}{4v^7}+\frac{9}{4 v^5}-\frac{3}{v^3}\right)a_1 a_{01}\right.\nn\\
&\left.-\left(\frac{1}{4v^5}-\frac{3}{4v^3}\right)\left(a_{01} \left(b_1+2 c_1\right)+4 a_{02}\right)\right]\nn\\
&+\left(1-v^2\right)^2\frac{a_{01}^3 \hat{q}^3 \left(1-v^2\right)^{3/2}}{2 v^7}.
\end{align}

In RN spacetime, the $y_3$ becomes
\begin{align}
y_3=& M^3\left\{ \frac{1}{2}\left(5+\frac{45}{ v^2}+\frac{15}{ v^4}-\frac{1}{ v^6}\right) \right.\nn\\
&+3 \left(-\frac{15}{2 v^2}-\frac{5}{v^4}+\frac{1}{2 v^6}\right)\hat{q}\hat{Q}\sqrt{1-v^2}\nn\\
&+3  \left[-\frac{1}{2}-\frac{3}{v^2}-\frac{1}{2 v^4}\right.\nn\\
&\left.+ \left(\frac{3}{2 v^2}+\frac{3}{v^4}-\frac{1}{2 v^6}\right)\hat{q}^2\left(1-v^2\right)\right]\hat{Q}^2 \nn\\
&+ \left[\left(\frac{9}{2 v^2}+\frac{3}{2 v^4}\right)\hat{q}\sqrt{1-v^2}\right.\nn\\
&\left.\left.+ \left(-\frac{3}{2 v^4}+\frac{1}{2 v^6}\right)\hat{q}^3\left(1-v^2\right)^{3/2}\right]\hat{Q}^3 \right\}
\label{eq:y3inrn}
\end{align}
It is seen that there is a term proportional to $\hat{q}\sqrt{1-v^2}\hat{Q}^3$. This indeed is a gravitational-electromagnetic coupling term, with a factor $\hat{q}\sqrt{1-v^2}\hat{Q}$ due to electromagnetic interaction and another factor $\hat{Q}^2$ due to gravitational contribution from the lens charge.


\begin{thebibliography}{100}

	\bibitem{Dyson:1920cwa} F.~W.~Dyson, A.~S.~Eddington and C.~Davidson,
  Phil.\ Trans.\ Roy.\ Soc.\ Lond.\ A {\bf 220}, 291 (1920).

	\bibitem{Sharon:2014ija} K.~Sharon and T.~L.~Johnson,
  Astrophys.\ J.\  {\bf 800}, no. 2, L26 (2015)

	\bibitem{Peng:2006ew} C.~Y.~Peng, C.~D.~Impey, H.~W.~Rix, C.~S.~Kochanek, C.~R.~Keeton, E.~E.~Falco, J.~Lehar and B.~A.~McLeod,
  Astrophys.\ J.\  {\bf 649}, 616 (2006)
  [astro-ph/0603248].

	\bibitem{Bartelmann:1999yn} M.~Bartelmann and P.~Schneider,
  Phys.\ Rept.\  {\bf 340}, 291 (2001)
  [astro-ph/9912508].

	\bibitem{Ade:2015zua} P.~A.~R.~Ade {\it et al.} [Planck Collaboration],
  Astron.\ Astrophys.\  {\bf 594}, A15 (2016)

	\bibitem{Refregier:2003ct} A.~Refregier,
  Ann.\ Rev.\ Astron.\ Astrophys.\  {\bf 41}, 645 (2003)
  [astro-ph/0307212].

	\bibitem{Lewis:2006fu} A.~Lewis and A.~Challinor,
  Phys.\ Rept.\  {\bf 429}, 1 (2006)
  [astro-ph/0601594].

	\bibitem{Metcalf:2001ap} R.~B.~Metcalf and P.~Madau,
  Astrophys.\ J.\  {\bf 563}, 9 (2001)
  [astro-ph/0108224].

	\bibitem{Hoekstra:2008db} H.~Hoekstra and B.~Jain,
  Ann.\ Rev.\ Nucl.\ Part.\ Sci.\  {\bf 58}, 99 (2008)

	\bibitem{Hirata:1987hu} K.~Hirata {\it et al.} [Kamiokande-II Collaboration],
Phys.\ Rev.\ Lett.\ {\bf 58}, 1490 (1987).

	\bibitem{Bionta:1987qt} R.~M.~Bionta {\it et al.},
Phys.\ Rev.\ Lett.\ {\bf 58}, 1494 (1987).

	\bibitem{IceCube:2018dnn} M.~G.~Aartsen {\it et al.} [IceCube and Fermi-LAT and MAGIC and AGILE and ASAS-SN and HAWC and H.E.S.S. and INTEGRAL and Kanata and Kiso and Kapteyn and Liverpool Telescope and Subaru and Swift NuSTAR and VERITAS and VLA/17B-403 Collaborations],
  Science {\bf 361}, no. 6398, eaat1378 (2018)

	\bibitem{IceCube:2018cha} M.~G.~Aartsen {\it et al.} [IceCube Collaboration],
  Science {\bf 361}, no. 6398, 147 (2018)

	\bibitem{Abbott:2016blz} B.~P.~Abbott {\it et al.} [LIGO Scientific and Virgo Collaborations],
  Phys.\ Rev.\ Lett.\  {\bf 116}, no. 6, 061102 (2016)
  doi:10.1103/PhysRevLett.116.061102
  [arXiv:1602.03837 [gr-qc]].

	\bibitem{Abbott:2016nmj} B.~P.~Abbott {\it et al.} [LIGO Scientific and Virgo Collaborations],
  Phys.\ Rev.\ Lett.\  {\bf 116}, no. 24, 241103 (2016)
  doi:10.1103/PhysRevLett.116.241103
  [arXiv:1606.04855 [gr-qc]].

	\bibitem{Abbott:2017oio} B.~P.~Abbott {\it et al.} [LIGO Scientific and Virgo Collaborations],
  Phys.\ Rev.\ Lett.\  {\bf 119}, no. 14, 141101 (2017)
  doi:10.1103/PhysRevLett.119.141101
  [arXiv:1709.09660 [gr-qc]].

	\bibitem{TheLIGOScientific:2017qsa} B.~P.~Abbott {\it et al.} [LIGO Scientific and Virgo Collaborations],
  Phys.\ Rev.\ Lett.\  {\bf 119}, no. 16, 161101 (2017)
  doi:10.1103/PhysRevLett.119.161101
  [arXiv:1710.05832 [gr-qc]].

	\bibitem{Monitor:2017mdv} B.~P.~Abbott {\it et al.} [LIGO Scientific and Virgo and Fermi-GBM and INTEGRAL Collaborations],
  Astrophys.\ J.\  {\bf 848}, no. 2, L13 (2017)
  doi:10.3847/2041-8213/aa920c
  [arXiv:1710.05834 [astro-ph.HE]].

	\bibitem{LetessierSelvon:2011dy} A.~Letessier-Selvon and T.~Stanev,
Rev. Mod. Phys. \textbf{83}, 907-942 (2011)
doi:10.1103/RevModPhys.83.907
[arXiv:1103.0031 [astro-ph.HE]].

	\bibitem{AlvesBatista:2019tlv} R.~Alves Batista, J.~Biteau, M.~Bustamante, K.~Dolag, R.~Engel, K.~Fang, K.~H.~Kampert, D.~Kostunin, M.~Mostafa and K.~Murase, \textit{et al.}
Front. Astron. Space Sci. \textbf{6}, 23 (2019)
doi:10.3389/fspas.2019.00023
[arXiv:1903.06714 [astro-ph.HE]].

	\bibitem{Jia:2015zon} X.~Liu, J.~Jia and N.~Yang,
Class. Quant. Grav. \textbf{33}, no.17, 175014 (2016)
doi:10.1088/0264-9381/33/17/175014
[arXiv:1512.04037 [gr-qc]].

\bibitem{He:2016vxc} 
  G.~He and W.~Lin,
  Class.\ Quant.\ Grav.\  {\bf 33}, no. 9, 095007 (2016)
  Addendum: [Class.\ Quant.\ Grav.\  {\bf 34}, no. 2, 029401 (2017)]
  doi:10.1088/0264-9381/33/9/095007, 10.1088/1361-6382/aa5203
  [arXiv:2007.08754 [gr-qc]].
  
	\bibitem{Pang:2018jpm} X.~Pang and J.~Jia,
Class. Quant. Grav. \textbf{36}, no.6, 065012 (2019)
doi:10.1088/1361-6382/ab0512
[arXiv:1806.04719 [gr-qc]].

	\bibitem{Huang:2020trl} K.~Huang and J.~Jia,
JCAP \textbf{08}, 016 (2020)
doi:10.1088/1475-7516/2020/08/016
[arXiv:2003.08250 [gr-qc]].

	\bibitem{Liu:2020wcu} H.~Liu and J.~Jia,
[arXiv:2006.03542 [gr-qc]].

	\bibitem{Liu:2020mkf} H.~Liu and J.~Jia,
Eur. Phys. J. C \textbf{80}, no.10, 932 (2020)
doi:10.1140/epjc/s10052-020-08496-5
[arXiv:2006.11125 [gr-qc]].

	\bibitem{Duan:2020tsq} Y.~Duan, W.~Hu, K.~Huang and J.~Jia,
Class. Quant. Grav. \textbf{37}, no.14, 145004 (2020)
doi:10.1088/1361-6382/ab852c
[arXiv:2001.03777 [gr-qc]].

	\bibitem{Jia:2020xbc} J.~Jia,
Eur. Phys. J. C \textbf{80}, no.3, 242 (2020)
doi:10.1140/epjc/s10052-020-7796-y
[arXiv:2001.02038 [gr-qc]].

\bibitem{He:2020eah} 
  G.~He, X.~Zhou, Z.~Feng, X.~Mu, H.~Wang, W.~Li, C.~Pan and W.~Lin,
  Eur.\ Phys.\ J.\ C {\bf 80}, no. 9, 835 (2020).
  doi:10.1140/epjc/s10052-020-8382-z
  
	\bibitem{Crisnejo:2018uyn} G.~Crisnejo and E.~Gallo,
Phys. Rev. D \textbf{97}, no.12, 124016 (2018)
doi:10.1103/PhysRevD.97.124016
[arXiv:1804.05473 [gr-qc]].

\bibitem{Jusufi:2018kry} 
  K.~Jusufi,
  Phys.\ Rev.\ D {\bf 98}, no. 6, 064017 (2018)
  doi:10.1103/PhysRevD.98.064017
  [arXiv:1806.01256 [gr-qc]].
  
	\bibitem{Jusufi:2019rcw} K.~Jusufi,
[arXiv:1906.12186 [gr-qc]].

\bibitem{Li:2019vhp} 
  Z.~Li, G.~He and T.~Zhou,
  Phys.\ Rev.\ D {\bf 101}, no. 4, 044001 (2020)
  doi:10.1103/PhysRevD.101.044001
  [arXiv:1908.01647 [gr-qc]].
  
	\bibitem{Crisnejo:2019ril} G.~Crisnejo, E.~Gallo and K.~Jusufi,
Phys. Rev. D \textbf{100}, no.10, 104045 (2019)
doi:10.1103/PhysRevD.100.104045
[arXiv:1910.02030 [gr-qc]].

\bibitem{Javed:2019ynm} 
  W.~Javed, R.~Babar and A.~Övgün,
  Phys.\ Rev.\ D {\bf 100}, no. 10, 104032 (2019)
  doi:10.1103/PhysRevD.100.104032
  [arXiv:1910.11697 [gr-qc]].
  
	\bibitem{Li:2019qyb} Z.~Li and J.~Jia,
Eur. Phys. J. C \textbf{80}, no.2, 157 (2020)
doi:10.1140/epjc/s10052-020-7665-8
[arXiv:1912.05194 [gr-qc]].

\bibitem{Li:2020dln} 
  Z.~Li and A.~Övgün,
  Phys.\ Rev.\ D {\bf 101}, no. 2, 024040 (2020)
  doi:10.1103/PhysRevD.101.024040, 10.20944/preprints201911.0195.v1
  [arXiv:2001.02074 [gr-qc]].
  
	\bibitem{Crisnejo:2019xtp} G.~Crisnejo, E.~Gallo and J.~R.~Villanueva,
Phys. Rev. D \textbf{100}, no.4, 044006 (2019)
doi:10.1103/PhysRevD.100.044006
[arXiv:1905.02125 [gr-qc]].

	\bibitem{Li:2020ozr} Z.~Li, Y.~Duan and J.~Jia,
[arXiv:2012.14226 [gr-qc]].

	\bibitem{Gibbons:1982ih} G.~W.~Gibbons,
Nucl. Phys. B \textbf{207}, 337-349 (1982)
doi:10.1016/0550-3213(82)90170-5

	\bibitem{Gibbons:1987ps} G.~W.~Gibbons and K.~i.~Maeda,
Nucl. Phys. B \textbf{298}, 741-775 (1988)
doi:10.1016/0550-3213(88)90006-5

	\bibitem{Garfinkle:1990qj} D.~Garfinkle, G.~T.~Horowitz and A.~Strominger,
Phys. Rev. D \textbf{43}, 3140 (1991)
[erratum: Phys. Rev. D \textbf{45}, 3888 (1992)]
doi:10.1103/PhysRevD.43.3140

	\bibitem{Cisterna:2014nua} A.~Cisterna and C.~Erices,
Phys. Rev. D \textbf{89}, 084038 (2014)
doi:10.1103/PhysRevD.89.084038
[arXiv:1401.4479 [gr-qc]].

	\bibitem{Rohrlich} F. Rohrlich, {\it Classical Charged Particles}, World Scientific, 2007, 3rd ediction

	\bibitem{Proceedings:2006uym} P.~Schneider, C.~S.~Kochanek and J.~Wambsganss,
Saas-Fee Advanced Courses \textbf{33}, pp.1-553 (2006)

	\bibitem{Anninos:2001yb} P.~Anninos and T.~Rothman,
  Phys.\ Rev.\ D {\bf 65}, 024003 (2002)
  doi:10.1103/PhysRevD.65.024003
  [gr-qc/0108082].

	\bibitem{Ivanov:2002jy} B.~V.~Ivanov,
  Phys.\ Rev.\ D {\bf 65}, 104001 (2002)
  doi:10.1103/PhysRevD.65.104001
  [gr-qc/0203070].

	\bibitem{Eiroa:2003wp} E.~F.~Eiroa and G.~E.~Romero,
  Gen.\ Rel.\ Grav.\  {\bf 36}, 651 (2004)
  doi:10.1023/B:GERG.0000016916.79221.24
  [gr-qc/0303093].

	\bibitem{Horvat:2008ch} D.~Horvat, S.~Ilijic and A.~Marunovic,
  Class.\ Quant.\ Grav.\  {\bf 26}, 025003 (2009)
  doi:10.1088/0264-9381/26/2/025003
  [arXiv:0807.2051 [gr-qc]].

	\bibitem{Maurya:2015wma} S.~K.~Maurya, Y.~K.~Gupta, S.~Ray and S.~R.~Chowdhury,
  Eur.\ Phys.\ J.\ C {\bf 75}, no. 8, 389 (2015).
  doi:10.1140/epjc/s10052-015-3615-2

	\bibitem{Morales:2018nmq} E.~Morales and F.~Tello-Ortiz,
  Eur.\ Phys.\ J.\ C {\bf 78}, no. 8, 618 (2018)
  doi:10.1140/epjc/s10052-018-6102-8
  [arXiv:1805.00592 [gr-qc]].

	\bibitem{Sharif:2018toc} M.~Sharif and S.~Sadiq,
  Eur.\ Phys.\ J.\ C {\bf 78}, no. 5, 410 (2018)
  doi:10.1140/epjc/s10052-018-5894-x
  [arXiv:1804.09616 [gr-qc]].
  
	\bibitem{Arbanil:2013pua} J.~D.~V.~Arba?il, J.~P.~S.~Lemos and V.~T.~Zanchin,
  Phys.\ Rev.\ D {\bf 88}, 084023 (2013)
  doi:10.1103/PhysRevD.88.084023
  [arXiv:1309.4470 [gr-qc]].

	\bibitem{bk:padamanabhan} T. Padmanabhan, {\it Gravitation: Foundations and Frontiers}, Cambridge University Press (2010), Eq. (2.94)

	\bibitem{eddington} Eddington, A. S., {\it The Internal Constitution of the Stars} (Cambridge University Press,  1926)

	\bibitem{bally} Bally J., Harrison E. R., Astrophysical J. {\bf 220}, 743 (1978).

	\bibitem{gldatabase} C.S. Kochanek, E.E. Falco, C. Impey, J. Lehar, B. McLeod, H.-W. Rix, 
https://lweb.cfa.harvard.edu/castles/

	\bibitem{Bhadra:2003zs} A.~Bhadra,
Phys. Rev. D \textbf{67}, 103009 (2003)
doi:10.1103/PhysRevD.67.103009
[arXiv:gr-qc/0306016 [gr-qc]].

	\bibitem{Keeton:2005jd} C.~R.~Keeton and A.~O.~Petters,
Phys. Rev. D \textbf{72}, 104006 (2005)
doi:10.1103/PhysRevD.72.104006
[arXiv:gr-qc/0511019 [gr-qc]].

	\bibitem{Ovgun:2018prw} A.~\"Ovg\"un, G.~Gyulchev and K.~Jusufi,
Annals Phys. \textbf{406}, 152-172 (2019)
doi:10.1016/j.aop.2019.04.007
[arXiv:1806.03719 [gr-qc]].

	\bibitem{Ong:2019glf} Y.~C.~Ong and Y.~Yao,
JHEP \textbf{10}, 129 (2019)
[erratum: JHEP \textbf{12}, 164 (2019)]
doi:10.1007/JHEP10(2019)129
[arXiv:1907.07490 [gr-qc]].

	\bibitem{Horowitz:1992ya} G.~T.~Horowitz,
[arXiv:gr-qc/9301008 [gr-qc]].

	\bibitem{Blaga:2014spa} C.~Blaga,
Serb. Astron. J. \textbf{190}, 41 (2015)
doi:10.2298/SAJ1590041B
[arXiv:1407.1504 [gr-qc]].

	\bibitem{Feng:2015wvb} X.~H.~Feng, H.~S.~Liu, H.~L\"u and C.~N.~Pope,
Phys. Rev. D \textbf{93}, no.4, 044030 (2016)
doi:10.1103/PhysRevD.93.044030
[arXiv:1512.02659 [hep-th]].

	\bibitem{Wang:2019cuf} C.~Y.~Wang, Y.~F.~Shen and Y.~Xie,
JCAP \textbf{04}, 022 (2019)
doi:10.1088/1475-7516/2019/04/022
[arXiv:1902.03789 [gr-qc]].

\end{thebibliography}
\end{document}